\documentclass[article, shortnames]{jss}

\author{\"{O}mer Nebil Yavero\u{g}lu\\ Imperial College London\And
        Sean M. Fitzhugh\\ University of California, Irvine\And
		Maciej Kurant\\ Google, Zurich\AND	
        Athina Markopoulou\\ University of California, Irvine \And
        Carter T. Butts\\ University of California, Irvine \AND
        Nata\v{s}a Pr\v{z}ulj\\Imperial College London}
\title{\pkg{ergm.graphlets}: A Package for ERG Modeling Based on Graphlet Statistics}

\Plainauthor{\"{O}mer Nebil Yavero\u{g}lu, Sean M. Fitzhugh, Maciej Kurant, Athina Markopoulou, Carter T. Butts, Nata\v{s}a Pr\v{z}ulj} %% comma-separated
\Plaintitle{ergm.graphlets: A Package for ERG Modeling Based on Graphlet Statistics} %% without formatting
\Abstract{
Exponential-family random graph models (ERGMs) are probabilistic network models that are parametrized by sufficient statistics based on structural (i.e., graph-theoretic) properties. The \pkg{ergm} package for the \proglang{R} statistical computing system is a collection of tools for the analysis of network data within an ERGM framework. Many different network properties can be employed as sufficient statistics for ERGMs by using the model terms defined in the \pkg{ergm} package; this functionality can be expanded by the creation of packages that code for additional network statistics. Here, our focus is on the addition of statistics based on graphlets. Graphlets are small, connected, and non-isomorphic induced subgraphs that describe the topological structure of a network. We introduce an \proglang{R} package called \pkg{ergm.graphlets} that enables the use of graphlet properties of a network within the \pkg{ergm} package of R. The \pkg{ergm.graphlets} package provides a complete list of model terms that allows to incorporate statistics of any 2-, 3-, 4- and 5-node graphlet into ERGMs. The new model terms of \pkg{ergm.graphlets} package enable both ERG modelling of global structural properties and investigation of relationships between nodal attributes (i.e., covariates) and local topologies around nodes.
}
\Keywords{graphlet, graphlet degree, subgraph, exponential-family random graph model,  \pkg{ergm}, \pkg{statnet}, \proglang{R}}
\Plainkeywords{graphlet, graphlet degree, subgraph, exponential-family random graph model, ergm, statnet, R} %% without formatting
\Address{
Nata\v{s}a Pr\v{z}ulj\\
Department of Computing\\   
Imperial College London\\
180 Queen's Gate\\    
LONDON, SW7 2AZ, UK\\
E-mail:  \email{natasha@doc.ic.ac.uk}\\
URL: \url{http://www.doc.ic.ac.uk/~natasha/}\\
Telephone: +44/207/594-8287\\
Fax: +44/207/594-8932
}
\begin{document}

%% include your article here, just as usual
%% Note that you should use the \pkg{}, \proglang{} and \code{} commands.

%% include your article here, just as usual
%% Note that you should use the \pkg{}, \proglang{} and \code{} commands.
\section[Introduction]{Introduction}
\label{sec:intro}
A \emph{graph} (or \emph{network}) is a representation of a set of objects and the relations among them. Each object is represented by a \emph{node} (or \emph{vertex}) in a graph. If there exists a relationship between two objects, their corresponding nodes are connected with an \emph{edge}. A graph is represented as $G=(V, E)$ where $V$ is the set of nodes and $E$ is the set of edges of graph $G$. The \emph{degree} of a node is the number of connections it has to the other nodes. The \emph{degree distribution} of a graph is the frequency distribution of the degrees of all nodes. Given two graphs, $G=(V, E)$ and $G'=(V', E')$, $G'$ is a \emph{subgraph} of $G$ if $V'$ is a subset of  $V$ and $E'$ is a subset of $E$. The subgraphs can have different topological structures; e.g., triangle, k-star, or k-cycle. A \emph{triangle} is a complete graph of order 3. A \emph{k-star} is defined to be a node $v$ and a set of $k$ distinct nodes, $\{v_1, \dots, v_k\}$, that are not equal to $v$ and $(v, v_i)\in E$ for each $i \in \{1, \dots, k\}$. A \emph{k-cycle} is defined to be a set of $k$ distinct nodes $\{v_1, v_2, \dots, v_k\}$, such that $(v_1, v_2)$, $(v_2, v_3)$, $\dots$, $(v_k, v_1) \in E$. If a subgraph $G'$ contains all edges in $G$ that connect nodes of $V'$, then $G'$ is said to be an \emph{induced} subgraph of $G$. An \emph{isomorphism}, $f$, from graph $G=(V_1, E_1)$ to graph $H=(V_2, E_2)$ is a bijection, $f: V_1 \rightarrow V_2$ , such that $(v_1, v_2)$ is an edge of $G$ if and only if $(f(v_1), f(v_2))$ is an edge of $H$. An \emph{automorphism} is an isomorphism from a graph to itself. The automorphisms of a graph $G$ under composition of functions form a group, called automorphism group of $G$, $Aut(G)$. The automorphism orbit of node $u$ is defined as $Orb(u) = \{y \in V(G) | y = f(x)$ for some $f \in Aut(G)\}$.  

\begin{figure}[t]  
  \centering
  \includegraphics[width=\textwidth]{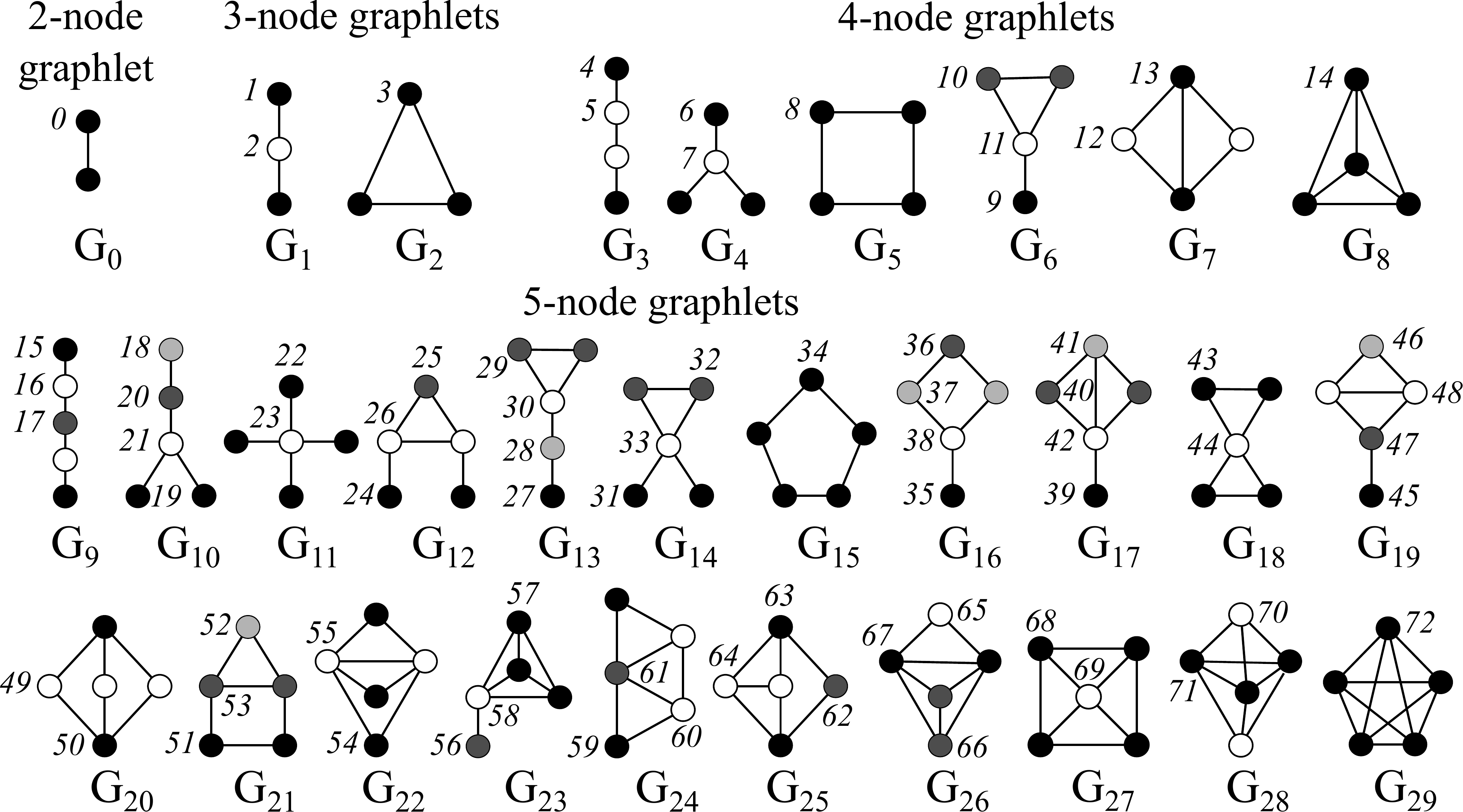}
  \caption{All 2-, 3-, 4- and 5-node graphlets, $G_{0}$, $G_{1}$, ..., $G_{29}$,  and their automorphism orbits, 0, 1, 2, ..., 72. \citep{graphlets2}}
  \label{fig:graphlets}
\end{figure}

\begin{figure}[t]  
  \centering
  \includegraphics[width=\textwidth]{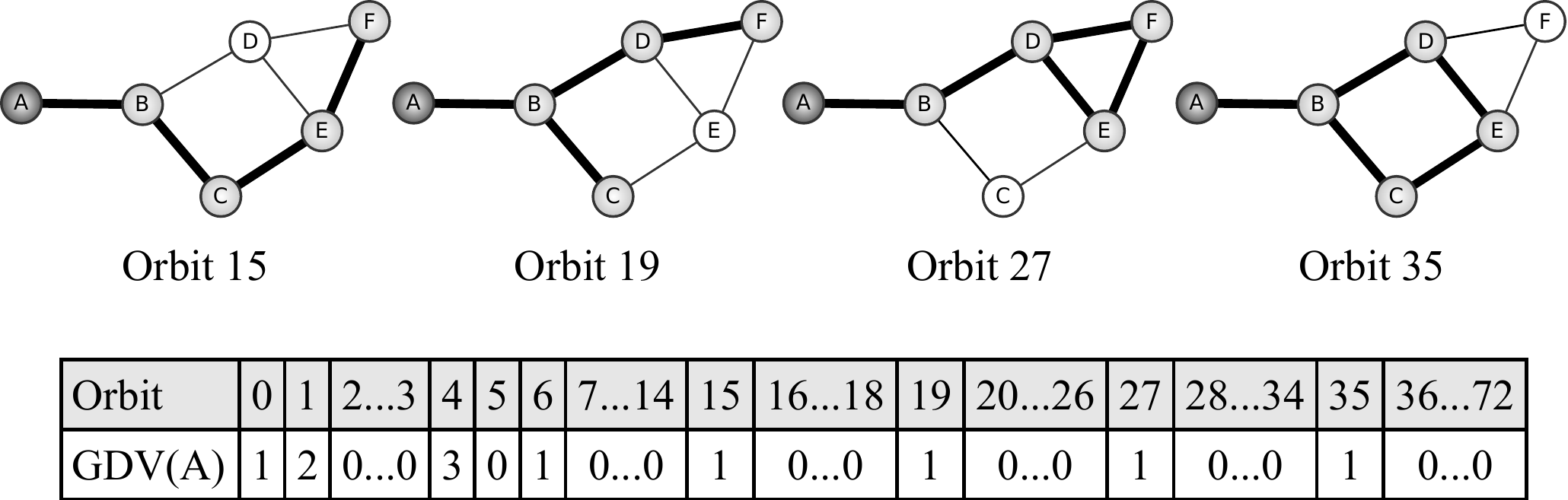}
  \caption{An illustration of the graphlet signature of node $A$. The number of graphlets that touch node $A$ at orbit $i$ is the $i^{th}$ element of graphlet degree vector of a node. The figure highlights the graphlets that touch node $A$ at orbits 15, 19, 27, and 35 from left to right, respectively \citep{graal}.}
  \label{fig:graphletCount}
\end{figure}

\emph{Graphlets} are small, connected, and non-isomorphic induced subgraphs of a graph that have  $n \geq 2$ nodes \citep{GEO1}. All graphlets with 2, 3, 4 and 5 nodes and their automorphism orbits are illustrated in Figure~\ref{fig:graphlets}. Graphlets have been used to describe the topological characteristics of a graph in two ways. The first technique uses the number of occurrences of each graphlet in the graph: the 30-dimensional vector obtained by counting the occurrences of each order $\le 5$ graphlet in the network is used for describing the topological structure of the graph \citep{GEO1}. The second technique employs the \emph{graphlet degree distribution} (GDD), which generalizes the degree distribution into the spectrum of 73 distributions, each corresponding to the distribution of the number of nodes touching a particular order $\le 5$ graphlet at a particular orbit \citep{graphlets2}. The graphlet degree distribution is used in a similar manner to the degree distribution to evaluate the topological characteristics of the whole network. It is a more refined measure of network topology than graphlet counts. The degree of a node can likewise be generalized into a 73-dimensional vector (again for order $\le 5$ graphlets) in which each coordinate represents the number of graphlets that the node touches at a particular orbit; this vector is called the \emph{graphlet degree vector} (GDV), or the \emph{graphlet signature} of the node \citep{graphletSignatures}. The computation of the graphlet signature of a node is illustrated in Figure~\ref{fig:graphletCount}. Graphlet-based methods have been successfully used for analyzing biological networks. There are various algorithms that use graphlets to evaluate topological similarities of two graphs \citep{graphlets2}, the fit of a graph to a network model \citep{GEO1,graphlets2}, to align two networks \citep{graal, HGRAAL, MIGRAAL, CGRAAL}, and to identify topologically similar nodes in a graph \citep{graphletSignatures, Milenkovic2009b}.

Understanding the processes underlying the formation of edges in a network is one of the main challenges in network modeling. Various network models describe  different rules for formation of edges; e.g., Erd\"{o}s-R\'{e}nyi random graph models (also known as Bernoulli graphs) \citep{ERmodels}, so-called ``scale-free'' models \citep{scalefree}, geometric models \citep{GEO2}, and stickiness-index-based models \citep{sticky}. Recent work on the statistical modeling of networks has focused on the use of discrete exponential families as general representations for these and other graph distributions. Exponential-family random graph models (ERG models or ERGMs, also known as ``p*'' models) are probabilistic network models that are parametrized in terms of sufficient statistics based on structural (i.e., graph-theoretic) properties \citep{holland.leinhardt:jasa:1981, exponential2, exponential3}. In ERGMs, the conditional probability of the existence of an edge given the rest of the graph in which it is determined by the effect that the edge has on the values of one or more network properties  (sufficient statistics or functions thereof), which are conventionally called model \emph{term}s.

The \pkg{ergm} package \citep{ergmPack, ergmpack11} for the \proglang{R} statistical computing system \citep{RTeam} provides a set of tools for analyzing networks within an ERGM framework. The \pkg{ergm} package allows the users to define ERGMs based on a wide range of network properties, fit ERGMs to observed networks using likelihood-based methods, simulate networks from an ERGM, perform graphical goodness-of-fit tests of the type described by \citet{modelChecking} and \citet{statnetManual}. The \pkg{ergm} package itself provides a large but limited number of model terms; this functionality can be extended by the creation of new packages which add additional terms \citep{ergm.userterms}. Due to \pkg{ergm}'s modular design, model terms added in this way may be employed with \pkg{ergm} in the precisely same manner as natively supported terms, and are transparent from an end-user perspective. Our focus in this paper is on the introduction of a new package of this type, \pkg{ergm.graphlets}, which adds support for model terms based on graphlet structures. The current subgraph pattern terms in \pkg{ergm} package (i.e., \textit{edges}, \textit{twopath}, \textit{triangle}, \textit{kstar}, \textit{cycle}) differ from the \pkg{ergm.graphlets} terms, as only induced subgraph patterns
are taken into account by the \pkg{ergm.graphlets} terms. \pkg{ergm.graphlets} package also provides an extended list of subgraph pattern based terms that covers any observable subgraph pattern of size 2, 3, 4, and 5.

\vspace{6 mm}
In the remainder of this article, we proceed as follows.  First, we provide a brief summary of ERG modeling \citep{bookSNA, ergmPack, ergmTermList, statnetManual} in order to make the article self-contained  (Section~\ref{sec:ergm}). Next, we provide installation details and licence information about the \pkg{ergm.graphlets} package together with the detailed explanations of the new model terms in Section~\ref{sec:terms}. The process of ERG modeling using the new terms of the \pkg{ergm.graphlets} package is illustrated in Section~\ref{sec:example}. The algorithms used for identifying the changes in graphlet statistics are described in Section~\ref{sec:algoritmic}. Section~\ref{sec:discussion} concludes the article with a brief summary and future directions.

\section[Exponential-family random graph modeling]{Exponential-family random graph modeling}
\label{sec:ergm}
Random graph models in ERGM form are specified via three elements: a vector of terms (sufficient statistics or functions thereof); a vector of real-valued parameters; and a support (often chosen to be the set of all graphs or digraphs of a given order) \citep{bookSNA, ergmPack}.\footnote{Technically, a \emph{reference measure} is also required; for unvalued graphs on finite support, this can be taken without loss of generality to be the counting measure.} Sufficient statistics for an ERGM can be functions representing any topological characteristics of the network (and, optionally, covariates), e.g., the number of edges, the degree distribution, the number of triangles, the number of k-stars, or the number of k-cycles. In general, few constraints on model terms are required; any real-valued functions are permissible, so long as they are finite and (for identifiable models) affinely independent on the support. Model terms can also relate nodal or edge attributes with their topological properties, e.g., the correlation between the node's attribute value and its degree. Readers can refer to \citet{ergmTermList} for a summary of model terms that are available in the \pkg{ergm} package.

Let $\mathbf{Y}$ be a random variable that represents an  $n$-by-$n$ adjacency matrix of an unweighted, loopless (no self-edges), undirected network with $n$ nodes.  $\mathbf{Y}$ can have $2^{n \choose 2}$ different values (configurations), where each value represents a different network having $n$ nodes. The number of configurations arises from the fact that there are $n \choose 2$ dyads in an order-$n$ graph, each of which may here take two distinct states. The set of all possible configurations is called the \textit{support} for $\mathbf{Y}$, denoted here by $\mathcal{Y}$. Any element of $\mathcal{Y}$ is a potential \emph{realization} of $\mathbf{Y}$ and is represented by $\mathbf{y}$.  An ERGM describes the probability of observing a realization, $\mathbf{y}$, as a function of a vector of sufficient statistics. The probability of observing a realization is expressed in ERGM form per Equation~\ref{eq:ergm}:
\begin{equation}
\Prob_{\theta, \mathcal{Y}} (\mathbf{Y} = \mathbf{y} | \theta , t) = \frac{\exp\{\theta^\top t(\mathbf{y})\}}{\sum_{z \in \mathcal{Y}} \exp\{ \theta^\top t(z)\}} , \mathbf{y} \in \mathcal{Y} ,
\label{eq:ergm}
\end{equation}
where $\theta$ is the vector of model coefficients (i.e., the weights for the model terms) and $t$ is the vector of sufficient statistics (i.e., model terms corresponding to network properties of interest) \citep{markovGraph1, markovGraph2}. Generalization of the above to more general cases (e.g., graphs with loops, digraphs, etc.) is immediate given alternative choice of $\mathcal{Y}$.  Since any probability mass function for $\mathbf{Y}$ on finite $\mathcal{Y}$ can be written in this form, ERGMs are a fully general representation for random graphs of finite order.

The denominator of Equation~\ref{eq:ergm} is a normalizing factor. The computation of the normalizing factor in the general case requires computation of the exponent term for all possible realizations of $\mathbf{Y}$, which typically has computational complexity of order $2^{n^2}$. In most cases, it is not possible to compute the normalizing factor in a reasonable time. However, for many purposes one can work with ratios of graph probabilities (i.e., $\Prob_{\theta,\mathcal{Y}}(\mathbf{Y}=\mathbf{y'}|\theta,t)/\Prob_{\theta,\mathcal{Y}}(\mathbf{Y}=\mathbf{y}|\theta,t)$) rather than with probabilities themselves, in which case the normalizing factor cancels and we are left with an expression in terms of the differences in model statistics under the respective graphs.  The vector $t(\mathbf{y'})-t(\mathbf{y})$ is known as the vector of change statistics for $\mathbf{y'}$ versus $\mathbf{y}$ under $t$, and plays a critical role in ERGM computation.  Of particular importance are the change statistics resulting from the perturbation of $\mathcal{Y}$ by a single edge state (i.e., adding or removing a specific edge). The change statistics under such a perturbation may be derived as follows. Let $\mathbf{y}$ be a realization of $\mathbf{Y}$. $\mathbf{y_{ij}^{+}}$ represents the configuration that contains all the edges of $\mathbf{y}$ and the edge between nodes $i$ and $j$. Similarly, $\mathbf{y_{ij}^{-}}$ represents the configuration that contains all the edges of $\mathbf{y}$ excluding the edge between nodes $i$ and $j$. Then, the change statistics of $\mathbf{y}$ for nodes $i$ and $j$ under perturbation of the $(i, j)$ edge,  $\delta_{t}(\mathbf{y})_{i,j}$, are defined as in Equation~\ref{eq:chStat2}:
\begin{equation}
\delta_{t}(\mathbf{y})_{i,j} = t(\mathbf{y}_{ij}^{+}) - t(\mathbf{y}_{ij}^{-}).
\label{eq:chStat2}
\end{equation}
Note that when considering only the relative probabilities of particular realizations, the normalizing factor can be eliminated by dividing $\Prob_{\theta, \mathcal{Y}} (\mathbf{Y} = \mathbf{y_{ij}^{+}} | \theta , t)$ by $\Prob_{\theta, \mathcal{Y}} (\mathbf{Y} = \mathbf{y_{ij}^{-}} | \theta , t)$. This leads to particularly simple form for the conditional odds for the existence of a given $(i, j)$ edge as in Equation~\ref{eq:logodds}:
\begin{equation}
\frac{\Prob_{\theta, \mathcal{Y}} (\mathbf{Y} = \mathbf{y_{ij}^{+}} | \theta , t)}{\Prob_{\theta, \mathcal{Y}} (\mathbf{Y} = \mathbf{y_{ij}^{-}} | \theta , t)} = \exp \{ \theta^\top \delta_{t}(\mathbf{y})_{i,j})\}.
\label{eq:logodds}
\end{equation}
The conditional odds given in Equation~\ref{eq:logodds} can be used for deriving the conditional probability of the existence of any single edge given the remainder of the graph.  Indeed, it follows immediately that the conditional edge probabilities are the inverse $\mathrm{logit}$s of the inner product of $\theta$ and the change statistics for the edge variable in question as in Equation~\ref{eq:edgeProb}:
\begin{equation}
\Prob_{\theta, \mathcal{Y}}(\mathbf{Y}_{ij} = 1 | \mathbf{Y}_{ij}^{c} = \mathbf{y}_{ij}^{c} , \theta, \delta_{t}) = \mathrm{logit}^{-1}(\theta^\top \delta_{t}(\mathbf{y})_{i,j}),
\label{eq:edgeProb}
\end{equation}
where $\mathrm{logit}(p) = \log[p / (1-p)]$,  and $\mathbf{y}_{ij}^{c}$ is the realization that contains all the edges of $\mathbf{y}$ except the edge ($i$, $j$).

In an inferential context, ERG models for an observed network, $y$, are typically fit by estimating the model coefficients, $\theta$, that maximize the conditional probability, $\Prob_{\theta, \mathcal{Y}} (Y=y | \theta , t)$ for some selected $t$ (with statistics being chosen based on a combination of exploratory analysis and prior theory). The most common approaches to estimation are currently Maximum Pseudo-Likelihood Estimation (MPLE, generally avoided except as an approximation) and Maximum Likelihood Estimation (MLE, implemented via one of several techniques). Current MLE methods (e.g., MCMC-MLE and stochastic approximation) typically rely on Markov chain Monte Carlo (MCMC) algorithms to simulate ERGM draws without computing normalizing factors. Although implementations vary, a typical MCMC algorithm for ERGM simulation randomly perturbs the edge  states in the simulated network one by one and uses the change statistics of these edge flips to compute the change in acceptance probabilities using Equations~\ref{eq:logodds}~or~\ref{eq:edgeProb}. For both estimation strategies, therefore, change statistics are employed as a means of avoiding explicit normalizing factor computation.  Indeed, one further consequence of this is that the model statistics themselves need never be directly computed: for most purposes, only the change scores are directly necessary.  Computing $\delta$ rather than $t$ can yield substantial savings for many commonly used model terms (e.g., degree statistics, k-stars, triad counts, etc.), a fact that is factored into the \pkg{ergm} package design.  Specifically, computation within \pkg{ergm} relies only on change statistics, and avoids direct computation of $t$.  Extension packages such as \pkg{ergm.graphlets}, therefore, must be written so as to rapidly compute change statistics (ideally without resorting to explicit calculation of the underlying graph statistics themselves.). The readers can refer to \citet{statnetTutorial} for details about using the \pkg{ergm} package. 

The \pkg{ergm.userterms} package \citep{ergm.userterms} enables users to define custom model terms for the \pkg{ergm} package. A new, user-defined term can be included into the \pkg{ergm} package by implementing a \proglang{R} function and a \proglang{C} function. The \proglang{R} function acts as an interface for the model term and preprocesses the term parameters before the computation of the change statistics. The \proglang{C} function performs the computation of the change statistics for the model term when an edge is flipped in the network. For example, for defining ``the number of edges'' term, the \proglang{C} function should return +1 when a new edge is added into the network and -1 when an edge is removed. The code for calculating the change statistics should be time-optimized, as it is likely that this computation will be performed millions of times during a typical MCMC run.

\section[The ergm.graphlets package]{The \pkg{ergm.graphlets} package}
\label{sec:terms}
The \pkg{ergm.graphlets} package is available from the Comprehensive R Archive Network (CRAN) at \url{http://CRAN.R-project.org/package=ergm.graphlets}. Since the \pkg{ergm.graphlets} package is an extension of the \pkg{ergm} package for \proglang{R}, users must first obtain the \proglang{R} framework (version 2.12 or above) and the \pkg{ergm} package (version 3.0.1 or above) before they can use the \pkg{ergm.graphlets} package \citep{ergmPack}. The \pkg{ergm.graphlets} package is open-source and released under GPL-2 and higher. Further information on the installation of the \pkg{ergm.graphlets} package and the licence can be obtained from the corresponding CRAN page.

The \pkg{ergm.graphlets} package introduces four graphlet based ERG modeling terms into the \pkg{ergm} package for \proglang{R}.  Graphlets identify the topological characteristics of a network associated with small induced subgraphs. Currently, \pkg{ergm.graphlets} provides support for sufficient statistics based on all graphlets of size 2, 3, 4 and 5; which are shown in Figure~\ref{fig:graphlets}.  The four model term functions that enable using graphlet statistics introduced by \pkg{ergm.graphlets} package are graphlet counts (\code{graphletCount}), graphlet orbit covariance (\code{grorbitCov}), graphlet orbit factor (\code{grorbitFactor}), and graphlet orbit distribution (\code{grorbitDist}). These terms can be used seamlessly with other terms from \pkg{ergm} or related packages. The functionality of the four new term functions is as follows:

\begin{enumerate}

\item{\textbf{Graphlet counts -- \code{graphletCount(g)}:}\\
Statistics for the number of times that a graphlet appears in a network can be included in an ERGM by using the \code{graphletCount} term. The question that the change score function of this term answers is: how do the number of graphlets of type $G_{i}$ change when an edge is flipped in the network? This term has an optional argument, $g$. $g$ is a vector of distinct integers representing the list of graphlets to be evaluated during the estimation of model coefficients (See Figure~\ref{fig:graphlets} for the list of graphlets). When this argument is not provided, all graphlets are evaluated by default. The term adds one network statistic to the model for each element in $g$. This term is defined for the 30 graphlets with up to 5 nodes. Therefore, $g$ accepts values between 0 and 29. Values outside this range are ignored. Some examples illustrating the usage of this term are provided in Section~\ref{sec:example}. The reader can also refer to Section~\ref{sec:algoritmic} for the  details about the implementation of the term.

The \code{graphletCount} term shows similarity with some terms of the \pkg{ergm} package, e.g., \code{cycle}, \code{edges}, \code{kstar}, \code{threepath}, \code{triangle}, \code{twopath}. There is a major difference between these terms and the \code{graphletCount} term. Graphlets are defined as induced subgraphs. Therefore, \code{graphletCount} does not count a subgraph as a two-path if the subgraph actually forms a triangle when induced on the nodes of the graph. In contrast, the terms in the \pkg{ergm} package do not require subgraphs to be induced. For this reason, a three node subgraph that forms a \code{triangle} is also counted as three \code{twopath} subgraphs by the \pkg{ergm} package. A closer parallel is the \code{triadcensus} term, which counts induced subgraphs on three nodes; note, however, that the triad census includes all isomorphism classes of order 3, while the order 3 graphlets consist only of the classes corresponding to connected graphs.  Thus, while there is overlap between some quantities computed by \code{graphletCount} and some existing \pkg{ergm} terms, the two are on the whole distinct.
}

\item{\textbf{Graphlet orbit covariance -- \code{grorbitCov(attrname, grorbit)}:}\\
The correlation between a node's graphlet degree and a numeric attribute value can be included into an ERGM by using the \code{grorbitCov} term. The question that the change score function of this term answers is: what is the change in covariance between a vector of nodal attributes and nodal orbit degrees when an edge is changed?  This term has two arguments: \code{attrname} and \code{grorbit}. The \code{attrname} is a character vector giving the name of a numeric attribute in the network's vertex attribute list. The optional \code{grorbit} argument is a vector of distinct integers representing the list of graphlet orbits to include into the ERGM model (See Figure~\ref{fig:graphlets} for the list of graphlet orbits). When \code{grorbit} is not provided, all graphlet orbits are evaluated by default. The term adds one network statistic to the model for each element in \code{grorbit}. Each term is equal to the sum given in Equation~\ref{eq:gorCov}:
\begin{equation}
grorbitCov(G, i, X) = \sum_{v \in V} GD_i(G,v) * X_v,
\label{eq:gorCov}
\end{equation}
where $X$ is the vector of attribute values, $i$ is the queried graphlet orbit and $GD_i(G,v)$ is the number of graphlets that touch node $v$ at orbit $i$. This term is defined for the 73 orbits corresponding to graphlets with up to 5 nodes. Therefore, \code{grorbit} accepts values between 0 and 72. Values outside this range are ignored. Some examples illustrating the usage of this term are provided in Section~\ref{sec:example}. The reader can also refer to Section~\ref{sec:algoritmic} for details about the implementation of the term. 

\code{grorbitCov} term can be viewed as extending the \code{nodecov} term in the \pkg{ergm} package to higher-order structures. In fact, \code{nodecov} term is a special case of \code{grorbitCov} where \code{grorbit} argument is set to 0.
}

\item{ \textbf{Graphlet orbit factor -- \code{grorbitFactor(attrname, grorbit, base)}:}\\
The \code{grorbitFactor} term adds a relationship between graphlet degrees and a categorical node attribute into an ERGM. The question that the change score function of this term answers is: what is the change in the total graphlet degree (for a given orbit) for those nodes with a given attribute value, for a particular edge change?  This term has three arguments: \code{attrname}; \code{grorbit}; and \code{base}.  \code{attrname} is a character vector giving the name of a categorical attribute in the network's vertex attribute list. The optional \code{grorbit} argument is a vector of distinct integers representing the list of graphlet orbits to include into the model (see Figure~\ref{fig:graphlets} for the list of graphlet orbits). When \code{grorbit} is not provided, all graphlet orbits are evaluated by default. The optional \code{base} argument is a vector of distinct integers representing the list of categories in \code{attrname} that are going to be omitted. When this argument is set to 0, all categories are evaluated. When this argument is set to $1$, the category having the lowest value (or lexicographically first name) is eliminated. The term sorts all values of the categorical attribute lexicographically and \code{base} term defines the indexes of the categories to be omitted in this sorted list. For example, if the ``fruit'' attribute has values ``orange'', ``apple'', ``banana'' and ``pear'', \code{grorbitFactor("fruit" , 0 , 2:3)} will ignore the ``banana'' and ``orange '' factors and evaluate the ``apple'' and ``pear'' factors. When the \code{base} argument is not provided, the argument is set to $1$ by default. The \code{grorbitFactor} term adds $a*|$\code{grorbit}$|$ terms into the model where $a$ represents the number of attribute values that are evaluated in the model and $|$\code{grorbit}$|$ is the number of graphlet orbits to be evaluated in the model. Each term is equal to the sum in Equation~\ref{eq:gorFactor}:
\begin{equation}
grorbitFactor(G, i, X_{c}) = \sum_{v \in V, category(v) = X_{c}}  GD_i(G,v),
\label{eq:gorFactor}
\end{equation}
where $X_{c}$ is the category of the term, $i$ is the queried graphlet orbit, $category(v)$ is the category that node $v$ belongs to, and $GD_i(G,v)$ is the number of graphlets that touch node $v$ at graphlet orbit $i$. This term is defined for the 73 graphlet orbits corresponding to graphlets with up to 5 nodes. Therefore, \code{grorbit} accepts values between 0 and 72. Values outside this range are ignored. Some examples illustrating the usage of this term are provided in Section~\ref{sec:example}. The reader can also refer to Section~\ref{sec:algoritmic} for details about the implementation of the term.
}

\code{grorbitFactor} term extends the \code{nodefactor} term in the \pkg{ergm} package. In fact, \code{nodefactor} term is a special case of \code{grorbitFactor} where \code{grorbit} argument is set to 0.

\item{\textbf{Graphlet degree distribution -- \code{grorbitDist(grorbit, d)}:}\\
The graphlet degree distributions of various graphlet orbits can be included into the ERGM by using the \code{grorbitDist} term. The question that the change score function of this term answers is: how do the number of nodes having graphlet degree $n$ for orbit $i$ change when an edge is flipped? This term has two arguments: \code{grorbit} and \code{d}. The \code{grorbit} argument is a vector of distinct integers representing the list of graphlet orbits to include into the model (See Figure~\ref{fig:graphlets} for the list of graphlet orbits). The \code{d} argument is a vector of distinct integers. This terms adds one network statistic to the model for each pairwise combination of the arguments in \code{grorbit} and \code{d} vectors. The statistic for the combination of (\textit{i}, \textit{j}) is equals to the number of nodes in the network that have graphlet degree \textit{j} for orbit \textit{i}. This term is defined for the 15 graphlet orbits corresponding to graphlets with up to 4 nodes. Therefore, \code{grorbit} accepts values between 0 and 14. Graphlets of size 5 are omitted for this term because of the high computational complexity of the change score computation of the term. Some examples illustrating the usage of this term are provided in Section~\ref{sec:example}. The reader can also find the details about the implementation of this term in Section~\ref{sec:algoritmic}.

The \code{grorbitDist} term extends the \code{degree} term in the \pkg{ergm} package. In fact, \code{degree} term is a special case of the \code{grorbitDist} where \code{grorbit} argument is set to 0. However, the \code{grorbitDist} function does not support the filtering functionalities of the \code{degree} term that are defined with the \code{by} and \code{homophily} arguments.
}

\end{enumerate}

\section[Illustration: ERGMs with graphlet terms]{Illustration: ERGMs with graphlet terms}
\label{sec:example}
In this section, we illustrate the use of terms from the \pkg{ergm.graphlets} package with two examples, one from the social sciences and one from the biological sciences.

\subsection{Lake Pomona emergent multi-organizational network (EMON)}

Our first example comes from \citet{Drabek1981}'s set of interorganizational communication networks in the context of search and rescue operations.  The setting for our example is the immediate aftermath of the capsizing of the Showboat Whippoorwill following its contact with a tornado near the southern shore of Lake Pomona, due south of Topeka, Kansas \citep{Drabek1981}.  Sixty passengers and crew were stranded in the lake, prompting the immediate response of the twenty organizations whose communication ties compose our network.

We use the graphlet terms to analyze patterns of brokerage in the organizational search and rescue network. Brokerage relations require (at least) three actors, one of whom bridges the connection between the two otherwise disconnected nodes (or sets of nodes, in extended brokerage structures)  \citep{GF1989,Marsden1982}. The broker has the opportunity to mediate and facilitate exchanges between two parties, where the units exchanged may be goods, services, information, or any other transferable entities.  Occupation of brokerage roles has been related to greater power in exchange networks \citep{Burt1992} and control of information in interorganizational disaster response networks \citep{MBB2012}.  Not all organizations are fit to occupy such roles, however, either by design or by happenstance \citep{MBB2012,LTBP2008}.  Previous studies of brokerage have been limited to the use of marginal tests to determine whether levels of brokerage exceed what we would expect by some baseline \citep{GF1989,MBB2012,LTBP2008,SAB2013}.  The \pkg{ergm.graphlets} package enables us to examine brokerage using conditional tests in which we can identify entities' propensities to occupy brokerage roles independent of confounding factors such as degree.  The \code{grOrbitFactor} and \code{grOrbitCov} terms allow us to determine whether occupation of local positions within graphlets is associated with particular covariates. These graphlet terms allow us to answer questions related to entities' local automorphism orbits (e.g., brokerage) in a model-based framework.

Before we proceed with modeling the network, we demonstrate how we construct the network used in our illustrative model.  Drabek's \emph{emon} dataset is found in the \pkg{network} package \citep{networkPack}, which is automatically loaded alongside the \pkg{ergm} package.  Although the network is represented as a digraph, the edge relations in the network are inherently undirected, as informants report on the \emph{existence} of communication ties.  We symmetrize the network via a union rule \citep{Krackhardt1987} to account for the undirected relations being measured.

\begin{CodeChunk}
\begin{CodeInput}
R> library(''statnet''); library(''ergm.graphlets'')
R> data(''emon'')			
R> emon.3 <- network(symmetrize(emon[[3]]), directed=F)  
\end{CodeInput}
\end{CodeChunk}

Having symmetrized the original network, we next add three vertex attributes used in the model.  Command rank score is each organization's rating of how strong of a role it has in the network's chain of command, as reported by other organizations participating in the search and rescue effort.  The location of each group's headquarters was also recorded; organizations were situated locally or non-locally in the Lake Pomona response.  Finally, we include the sponsorship level of each organization: city, county, state, federal, or private.  

\begin{CodeChunk}
\begin{CodeInput}
R> Command.Rank.Score <- emon[[3]] %v% "Command.Rank.Score"
R> Location <- emon[[3]] %v% "Location"
R> Sponsorship <- emon[[3]] %v% "Sponsorship"
\end{CodeInput}
\end{CodeChunk}

When ranking those with the strongest role in the chain of command, informants were limited to the six organizations present from the early phase of the response.  As a result, some organizations were not ranked and have been coded ``NA'' in the \emph{emon} data.  For our example, we assume those who were not ranked have the lowest possible command rank score (being more marginal to the unfolding response) and assign them a score of 0.

\begin{CodeChunk}
\begin{CodeInput}
R> Command.Rank.Score[is.na(Command.Rank.Score)] <- 0
\end{CodeInput}
\end{CodeChunk}

We assign the vertex attributes from the original network to our symmetrized, undirected network.

\begin{CodeChunk}
\begin{CodeInput}
R> emon.3 %v% "Command.Rank.Score" <- Command.Rank.Score
R> emon.3 %v% "Location" <- Location 
R> emon.3 %v% "Sponsorship" <- Sponsorship
\end{CodeInput}
\end{CodeChunk}

\begin{figure}[!h]
  \centering
  \includegraphics[width=.5\textwidth]{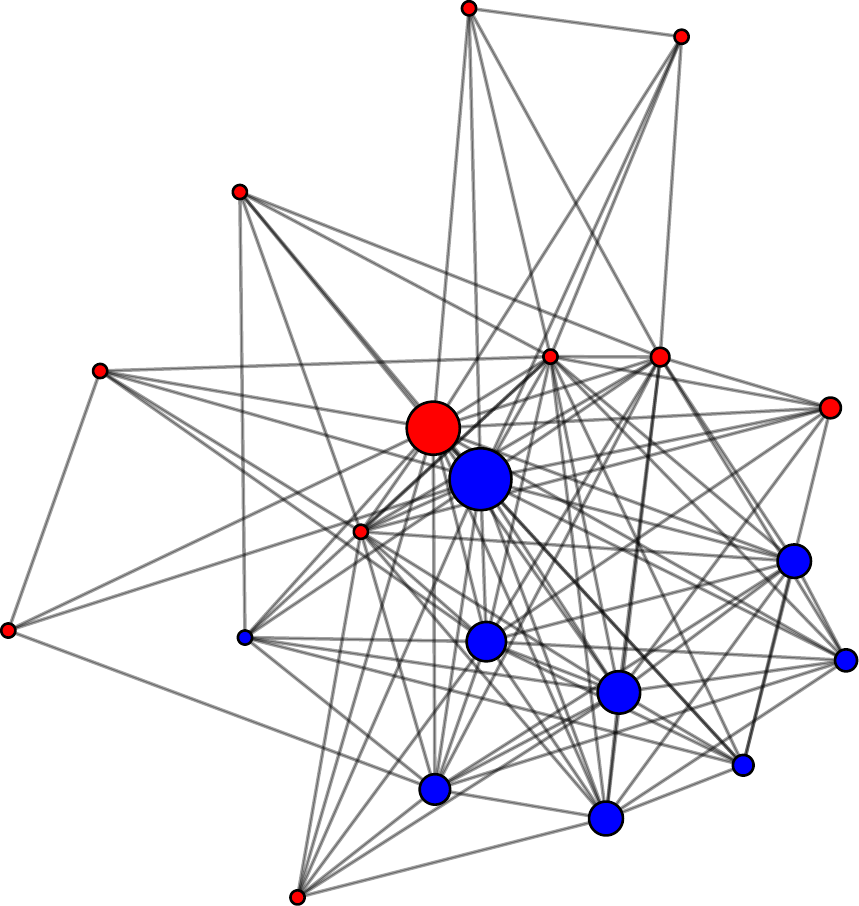}
  \caption{Lake Pomona emergent multi-organizational network (EMON) tasked with a search and rescue operation.  Node size is scaled to command rank score and nodes are colored by whether they had permanent headquarters situated locally (red) or non-locally (blue).}
  \label{f_EMONnet}
\end{figure}

We illustrate the network in Figure~\ref{f_EMONnet}. Our network resembles a core-periphery structure with the core primarily composed of non-local organizations and organizations with high command rank scores.

We are now ready to model the network with our graphlet terms.  We use an edge term for the total number of edges.  We use dyadic independence terms for sponsorship level and command rank score.  One might expect organizations at different sponsorship levels to be involved with more or fewer communication partnerships than organizations from a different sponsorship;  likewise, an organization's command rank score may be associated with its propensity to be involved in more communication partnerships. Finally, we include terms related to graphlet structure.  Graphlet $G_6$, which involves brokerage between cliques and individual nodes, is a natural choice given the core-periphery structure of the graph, and we include all its automorphism orbits---9, 10, and 11---into our model.  We incorporate the location covariate into the term to evaluate whether an organization's location is associated with its propensity to occupy these specific automorphism orbits.  The results will demonstrate whether the location of an organization in this type of subgraph is associated with its role as a pendant (automorphism 9), member of a dyad with ties to a broker (automorphism 10), or broker between the pendant and the dyad (automorphism 11).  We model the network below:

\begin{CodeChunk}
\begin{CodeInput}
R> emon.ergm <- ergm(emon.3 ~ edges + nodefactor("Sponsorship") +
+ nodecov("Command.Rank.Score") + 
+ grorbitFactor("Location", c(9:11)),
+ control =
+ control.ergm(seed = 1, MCMC.samplesize = 50000,
+ MCMC.interval = 100000, MCMC.burnin = 50000, parallel = 60))
\end{CodeInput}
\begin{CodeOutput}
Iteration 1 of at most 20: 
Loading required package: rlecuyer
Convergence test P-value: 0e+00 
The log-likelihood improved by 0.6982 
Iteration 2 of at most 20: 
Convergence test P-value: 0e+00 
The log-likelihood improved by 0.1079 
Iteration 3 of at most 20: 
Convergence test P-value: 0e+00 
The log-likelihood improved by 0.03083 
Iteration 4 of at most 20: 
Convergence test P-value: 0e+00 
The log-likelihood improved by 0.009207 
Iteration 5 of at most 20: 
Convergence test P-value: 2.2e-219 
The log-likelihood improved by 0.002616 
Iteration 6 of at most 20: 
Convergence test P-value: 4.8e-51 
The log-likelihood improved by 0.000652 
Iteration 7 of at most 20: 
Convergence test P-value: 2e-09 
The log-likelihood improved by 0.0001452 
Iteration 8 of at most 20: 
Convergence test P-value: 2.6e-03 
The log-likelihood improved by < 0.0001 
Iteration 9 of at most 20: 
Convergence test P-value: 9e-01 
Convergence detected. Stopping.
The log-likelihood improved by < 0.0001 

This model was fit using MCMC.  To examine model diagnostics and check
for degeneracy, use the mcmc.diagnostics() function.
\end{CodeOutput}
\end{CodeChunk}

Before examining the coefficients we examine the MCMC diagnostics to ensure the estimation process did not exhibit any peculiar behavior \citep{modelChecking}.  This model appears to have converged properly.

A summary of the model object reproduces the original formula for the model, the coefficients, deviance measures, and measures of the goodness of fit.

\begin{CodeChunk}
\begin{CodeInput}
R> summary(emon.ergm)
\end{CodeInput}
\begin{CodeOutput}
==========================
Summary of model fit
==========================

Formula:   emon.3 ~ edges + nodefactor("Sponsorship") + 
nodecov("Command.Rank.Score") + grorbitFactor("Location", c(9:11))

Iterations:  20

Monte Carlo MLE Results:
                                Estimate Std. Error MCMC %  p-value
edges                          -2.450670   0.688351      9 0.000473 ***
nodefactor.Sponsorship.County  -0.437354   0.319080      3 0.172175
nodefactor.Sponsorship.Federal -0.581708   0.606596      5 0.338852
nodefactor.Sponsorship.Private -0.041876   0.188267      1 0.824230
nodefactor.Sponsorship.State   -1.326516   0.785447      1 0.092967 .
nodecov.Command.Rank.Score      0.333315   0.075229      5  < 1e-04 ***
grorbitFactor.orb_9.attr_NL     0.009319   0.020540      0 0.650596
grorbitFactor.orb_10.attr_NL   -0.018051   0.014288      2 0.208081
grorbitFactor.orb_11.attr_NL    0.158800   0.031310      7  < 1e-04 ***
---

Signif. codes:  0 '***' 0.001 '**' 0.01 '*' 0.05 '.' 0.1 ' ' 1

     Null Deviance: 263.4  on 190  degrees of freedom
 Residual Deviance: 144.8  on 181  degrees of freedom

AIC: 162.8    BIC: 192    (Smaller is better.)
\end{CodeOutput}
\end{CodeChunk}

The results show significant effects for our \proglang{edge} term, command rank score, and non-local organizations' occupation of orbit 11.  The results show a strong, positive association between an organization's command rank score and its odds of forming a tie.  Most relevant to our interests, we find that one of the automorphism orbit terms is significant.  We find a positive, significant association between non-local (NL) organizations and their propensity to occupy a brokerage role between a pendant and a dyad (automorphism 11).  Substantively, this demonstrates that non-local organizations tend to occupy this specific structure of extended brokerage in which an organization occupies a brokerage position between one organization and a pair of connected organizations.

We use the \code{gof} command to examine the goodness of fit.  While the AIC and BIC demonstrate substantial improvements over a baseline model, the \code{gof} command measures demonstrate how well networks simulated from our model reproduce statistics from the original network.  We examine the model's reproduction of four statistics: geodesic distance, degree distribution, edgewise shared partner distribution, and the triad census.  We demonstrate below how we produce plots to examine these measures of fit.

\begin{CodeChunk}
\begin{CodeInput}
R> EMONgof <- gof(emon.ergm, GOF = ~ degree + distance + espartners 
		+ triadcensus)
R> par(mfrow = c(2, 2))
R> plot(EMONgof)
\end{CodeInput}
\end{CodeChunk}

\begin{figure}[!h]
  \centering
  \includegraphics{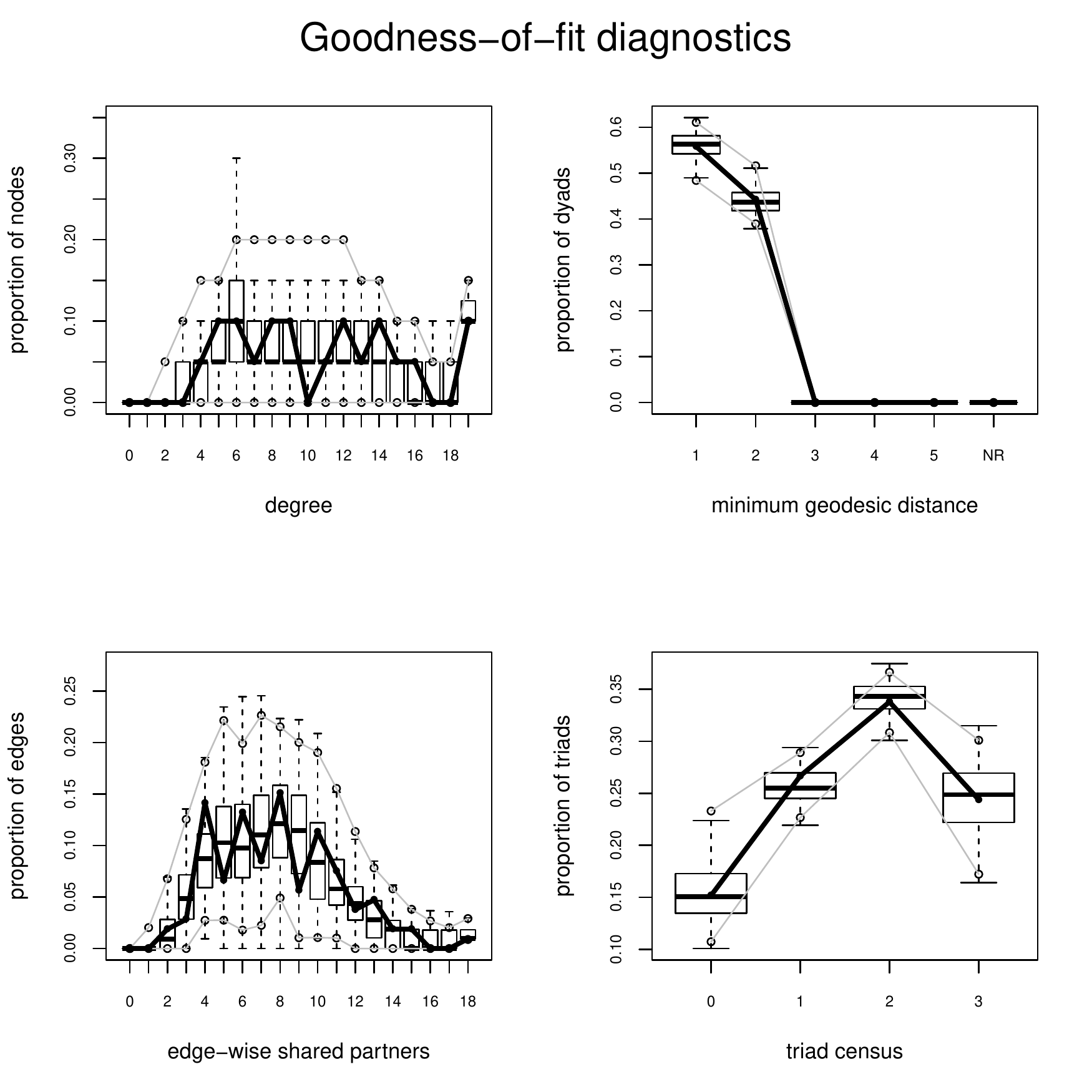}
  \caption{The solid black line in each plot represents the Lake Pomona EMON's observed statistics.  The boxplots illustrate the statistics for our simulated networks, as produced by the MLE.}
  \label{f_EMONgof}
\end{figure}

The plots are illustrated in Figure~\ref{f_EMONgof}.

As there are no clear discrepancies between the model-simulated networks and the original network, we find the model to be an adequate fit.

The graphlet orbit terms enable us to link local position to covariates in a model-based framework.  As demonstrated, this is a useful tool for modeling brokerage as we are able to link an entity's covariates to its propensity to occupy a specific brokerage role, whether it is a traditional (i.e., \code{twopath}) brokerage role or an extended brokerage role (e.g., orbit 11 in our model).  Beyond brokerage, these techniques can extend to any particular automorphism orbit contained within a graphlet:  pendants, clique members, or other nodes whose position may be linked to some categorical or continuous variable.  Being able to incorporate these covariate-driven graphlet terms into a model-based framework will enhance our ability to understand which factors are associated with nodes' occupation of local positions within graphlets.

%%%%%%%%%%%%%%%%%%%%%%%%%%%%%%%%%%%%%%%%%%%%%%%%%%%%%%%%%%%%%%%%%%%%%%%%%%%%%%%%%%%%%%%%%%%%%%%%%%%%%%%
%%%%%%%%%%%%%%%%%%%%%%%%%%%%%%%%%%%%%%%%%%%%%%%%%%%%%%%%%%%%%%%%%%%%%%%%%%%%%%%%%%%%%%%%%%%%%%%%%%%%%%%
%%%%%%%%%%%%%%%%%%%%%%%%%%%%%%%%%%%%%%%%%%%%%%%%%%%%%%%%%%%%%%%%%%%%%%%%%%%%%%%%%%%%%%%%%%%%%%%%%%%%%%%
%%%%%%%%%%%%%%%%%%%%%%%%%%%             PROTEINS                   %%%%%%%%%%%%%%%%%%%%%%%%%%%%%%%%%%%%
%%%%%%%%%%%%%%%%%%%%%%%%%%%%%%%%%%%%%%%%%%%%%%%%%%%%%%%%%%%%%%%%%%%%%%%%%%%%%%%%%%%%%%%%%%%%%%%%%%%%%%%
%%%%%%%%%%%%%%%%%%%%%%%%%%%%%%%%%%%%%%%%%%%%%%%%%%%%%%%%%%%%%%%%%%%%%%%%%%%%%%%%%%%%%%%%%%%%%%%%%%%%%%%
%%%%%%%%%%%%%%%%%%%%%%%%%%%%%%%%%%%%%%%%%%%%%%%%%%%%%%%%%%%%%%%%%%%%%%%%%%%%%%%%%%%%%%%%%%%%%%%%%%%%%%%

\subsection{Protein secondary structure network}

We next illustrate the graphlet count terms with a biological example.  The past decade has seen a surge of interest in identifying network motifs, or induced subgraphs whose size often ranges three to five nodes.  Applications span a wide variety of networks including transcription networks \citep{MiloScience2002,MiloScience2004,YegerLotemPNAS2004,ShenOrrNatureGenetics2002,Alon2007}, neuron synaptic connection networks \citep{MiloScience2002,MiloScience2004,KA2005}, protein-protein interaction networks \citep{AA2004,MiloScience2004,YegerLotemPNAS2004,Alon2007}, circuitry networks \citep{MiloScience2002,MiloScience2004,KA2005}, worldwide web networks \citep{MiloScience2002,MiloScience2004}, language networks \citep{MiloScience2004}, and social networks \citep{MiloScience2004}.  Typically scholars have used marginal tests to identify how frequently these subgraphs occur relative to some baseline.  In these types of tests the observed network is compared a set of randomized networks that hold constant some statistic of the original network, often the degree distribution.  While these types of marginal tests have been employed by networks scholars for decades \citep[see, e.g.,][for reviews]{WF1994,Butts2008}, a model-based approach allows us to examine the likelihood of observing these graphlets, conditioned on a variety of parameters (e.g., degree, triadic closure, covariates, etc.).  This is particularly important where the method of data collection itself may bias structure in particular ways; failure to account for these effects may result in spurious findings.  Below we use the \code{graphletCount} terms to examine patterns of biological network motifs in an ERGM framework, while controlling for artefacts of the data collection process.

Our example network is a protein structure network whose nodes are secondary structure elements (specifically, $\alpha$ helices and $\beta$ sheets) which are tied if the distance between them is smaller than 10 Angstroms (\r{A}) \citep{MiloScience2004} \footnote{This protein structure network can be obtained from: \url{http://www.weizmann.ac.il/mcb/UriAlon/Papers/networkMotifs/1eawInter_st.txt}}. This network represents the proximity structure of a matriptase-aprotinin complex (protein databse ID 1eaw) \citep{FriedrichPDB2002} as determined by x-ray crystallography (resolution 2.93\r{A}).  \citet{MiloScience2004} examine the overrepresentation and underrepresentation of graphlets in this network, by comparison to uniform random graphs conditional on the degree distribution.  They find that graphlets $G_3$ and $G_4$ are underrepresented while graphlets $G_6$, $G_7$, and $G_8$ are overrepresented.  We will determine whether these results hold in a model-based framework that allows us to account for potentially confounding degree, transitivity, and mixing effects, some of which represent artefacts of the data collection process.

We begin by reading in the data and converting it from an edge list into a \pkg{network} object.

\begin{CodeChunk}
\begin{CodeInput}
R> spi <- read.table("serine protease inhibitor.txt")
R> spi <- as.matrix(spi) 
R> attr(spi, "n") <- 53
R> spi <- as.sociomatrix.sna(spi)
R> spi <- network(spi, directed=F)
\end{CodeInput}
\end{CodeChunk}

\begin{figure}[!h]
  \centering
  \includegraphics[width=.5\textwidth]{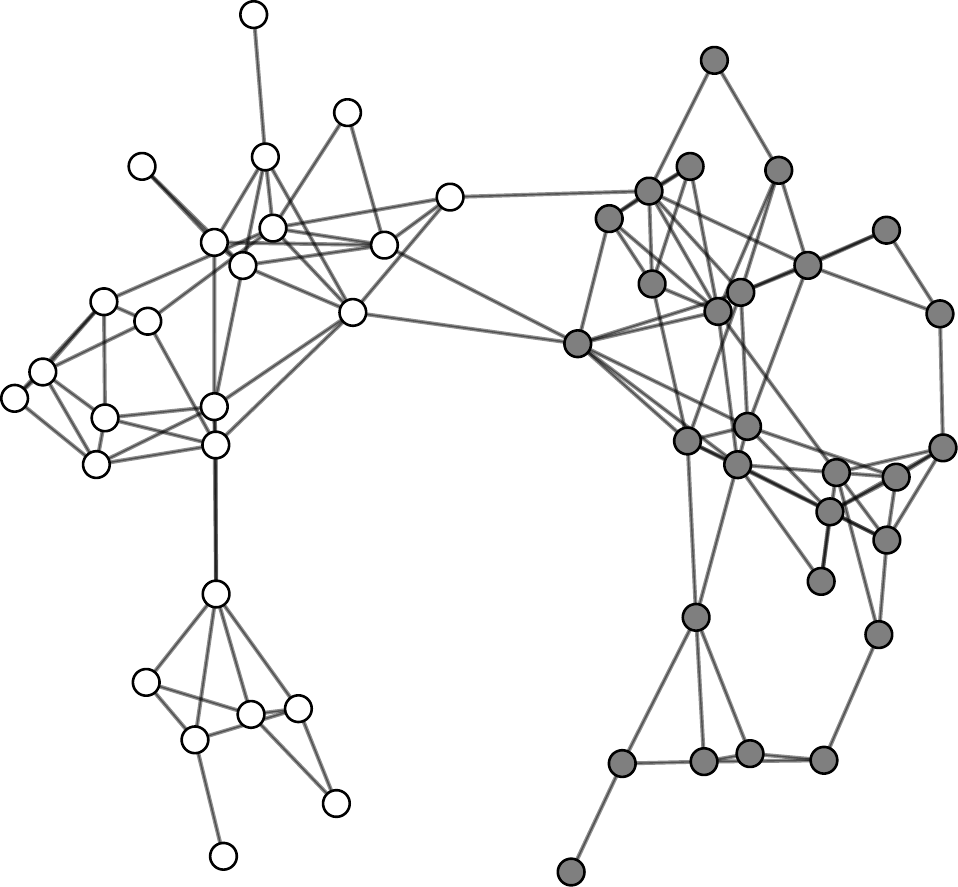}
  \caption{Network representation of the protein structure of the two matriptase-BPTI complexes.  Secondary structure elements are shaded by the complex to which they belong.}
  \label{f_SPInet} 
\end{figure}

Before modeling the protein structure network, it is important to consider how this network was obtained.  Although \citet{MiloScience2004} do not report on the content of the structure\footnote{The structure is not described in the paper, and is (inaccurately) summarized in the supplemental materials only as ``a serine protease inhibitor'' (Table S1).  In fact, the structure contains two assemblies, each of which is a complex of one domain of a serine protease (MT-SP1) with an inhibitor (BPTI).}, \citet{FriedrichPDB2002} note that the asymmetric unit of the crystal structure (from which the network is taken) contains two biological assemblies, each of which is a complex of two proteins (the catalytic domain of matriptase/MT-SP1 and a bovine pancreatic trypsin inhibitor/BPTI).  The presence of multiple copies of a biologically relevant complex within a crystal structure is a common artefact of the crystallization process, and indeed the same system could potentially have been observed with more or fewer complexes in the asymmetric unit.  This is of considerable importance for modeling the resulting network, as we would typically expect far more adjacencies within complexes than between them; failure to control for this effect may lead to very misleading conclusions.  Indeed, as shown in Figure~\ref{f_SPInet}, the network is dominated by two dense subgraphs corresponding to the two complexes, with very few ties spanning these subgraphs.  To account for this, we create vertex attributes based on biological assembly membership as reconstructed from information in the Protein Data Bank \citep{FriedrichPDB2002}, with polypeptide chains A and B of the structure belonging to assembly 1, and chains C and D belonging to assembly 2.  By incorporating these attributes into the model, we are much better able to account for the patterns of clustering in the network than we would be if we neglected the data collection process.   

We create an object to represent assembly membership and assign vertex attributes accordingly.

\begin{CodeChunk}
\begin{CodeInput}
R> spi.assembly <- 1:53
R> spi.assembly[1:25] <- "1"
R> spi.assembly[26:53] <- "2"
R> spi %v% "Assembly" <- spi.assembly
\end{CodeInput}
\end{CodeChunk}

We are now ready to model our network.  We begin by setting up our model with an edge term, a dyadic independence term, and several dyadic dependence terms, including our graphlet terms.  Because we observe very little tie formation across the sets of chains associated with each complex, we include a homophily term for protein assembly in our model.  Additionally, we include a within-assembly triadic closure term (i.e., closure of triads where all members belong to the same assembly).  We also include a degree term as the original paper was concerned with graphlet counts net of the degree distribution.  Of principal interest is our \code{graphletCount} term, which includes graphlets $G_3$, $G_4$, $G_6$, $G_7$, and $G_8$, the same set \citet{MiloScience2004} finds to occur at greater or lesser levels than chance.

Our first model includes all terms described above.  To speed up model fit, one may omit the ``control'' arguments, although the resulting standard errors (and accordingly, $p$~values) will be larger than what we report.

\begin{CodeChunk}
\begin{CodeInput}
R> spi.ergm.34678 <- ergm(spi ~ edges + nodematch("Assembly") + 
+ triangle("Assembly") + gwdegree(.5, fixed=T) + 
+ graphletCount(c(3, 4, 6, 7, 8)),
+ control =
+ control.ergm(seed = 1, MCMC.samplesize = 500000,
+ MCMC.interval = 75000, MCMC.burnin = 300000, parallel = 60))
\end{CodeInput}
\begin{CodeOutput}
Iteration 1 of at most 20: 
Loading required package: rlecuyer
Convergence test P-value: 0e+00 
The log-likelihood improved by 0.3676 
Iteration 2 of at most 20: 
Convergence test P-value: 0e+00 
The log-likelihood improved by 0.06913 
Iteration 3 of at most 20: 
Convergence test P-value: 0e+00 
The log-likelihood improved by 0.01911 
Iteration 4 of at most 20: 
Convergence test P-value: 0e+00 
The log-likelihood improved by 0.005275 
Iteration 5 of at most 20: 
Convergence test P-value: 0e+00 
The log-likelihood improved by 0.00151 
Iteration 6 of at most 20: 
Convergence test P-value: 0e+00 
The log-likelihood improved by 0.0004391 
Iteration 7 of at most 20: 
Convergence test P-value: 1.7e-83 
The log-likelihood improved by 0.0001026 
Iteration 8 of at most 20: 
Convergence test P-value: 5.2e-17 
The log-likelihood improved by < 0.0001 
Iteration 9 of at most 20: 
Convergence test P-value: 2.4e-03 
The log-likelihood improved by < 0.0001 
Iteration 10 of at most 20: 
Convergence test P-value: 8.3e-01 
Convergence detected. Stopping.
The log-likelihood improved by < 0.0001 

This model was fit using MCMC.  To examine model diagnostics and
check for degeneracy, use the mcmc.diagnostics() function.
\end{CodeOutput} 
 
\begin{CodeInput}
R> summary(spi.ergm.34678)
\end{CodeInput}
\begin{CodeOutput}
==========================
Summary of model fit
==========================

Formula:   spi ~ edges + nodematch("Assembly") + triangle("Assembly") + 
gwdegree(0.5, fixed = T) + graphletCount(c(3, 4, 6, 7, 8))

Iterations:  20 

Monte Carlo MLE Results:
                   Estimate Std. Error MCMC %  p-value    
edges              -6.42760    1.22926     12  < 1e-04 ***
nodematch.Assembly  2.48031    0.74204      6 0.000852 ***
triangle.Assembly   3.87343    0.67331      1  < 1e-04 ***
gwdegree            2.40227    1.51019      5 0.111906    
graphlet.3.Count    0.04962    0.02964      7 0.094298 .  
graphlet.4.Count   -0.03917    0.05467      1 0.473841    
graphlet.6.Count   -0.15361    0.04993      0 0.002137 ** 
graphlet.7.Count   -0.47295    0.17782      0 0.007910 ** 
graphlet.8.Count   -2.49869    0.72543      0 0.000590 ***
---

Signif. codes:  0 '***' 0.001 '**' 0.01 '*' 0.05 '.' 0.1 ' ' 1

     Null Deviance: 1910.3  on 1378  degrees of freedom
 Residual Deviance:  593.9  on 1369  degrees of freedom
 
AIC: 611.9    BIC: 658.9    (Smaller is better.) 
\end{CodeOutput}
\end{CodeChunk}

Our model finds a significant, positive effect for within-assembly homophily, a positive effect for triadic closure within complexes, and a propensity for the graph to be biased against formation of graphlets $G_6$, $G_7$, and $G_8$, assuming all other terms are held constant.  We find no significant results for graphlets $G_3$ and $G_4$.

We proceed to remove the non-significant terms to see if that improves model fit.  AIC suffers slightly if we remove $G_3$ from the model (AIC: 612.97), while BIC improves (654.8).  Both improve if we keep $G_3$ and remove $G_4$ (AIC: 610.73, BIC: 652.56).  We find the best fit by removing both $G_3$ and $G_4$ (AIC: 610.7, BIC: 647.3).  Accordingly, we fit our final model as follows.

\begin{CodeChunk}
\begin{CodeInput}

R> spi.ergm.all <- ergm(spi ~ edges + nodematch("Assembly") +
+ triangle("Assembly") + gwdegree(.5, fixed=T) +
+ graphletCount(c(6, 7, 8)),
+ control =
+ control.ergm(seed = 1, MCMC.samplesize = 15000,
+ MCMC.interval = 2000, MCMC.burnin = 15000))
\end{CodeInput}
\begin{CodeOutput}
Iteration 1 of at most 20:
Convergence test P-value: 0e+00
The log-likelihood improved by 0.2026
Iteration 2 of at most 20:
Convergence test P-value: 0e+00
The log-likelihood improved by 0.05503
Iteration 3 of at most 20:
Convergence test P-value: 2.5e-311
The log-likelihood improved by 0.01452
Iteration 4 of at most 20:
Convergence test P-value: 1.1e-86
The log-likelihood improved by 0.004337
Iteration 5 of at most 20:
Convergence test P-value: 1.6e-23
The log-likelihood improved by 0.001283
Iteration 6 of at most 20:
Convergence test P-value: 7.7e-04
The log-likelihood improved by 0.0002488
Iteration 7 of at most 20:
Convergence test P-value: 9.2e-03
The log-likelihood improved by 0.0001711
Iteration 8 of at most 20:
Convergence test P-value: 9.7e-01
Convergence detected. Stopping.
The log-likelihood improved by < 0.0001

This model was fit using MCMC.  To examine model diagnostics
and check for degeneracy, use the mcmc.diagnostics() function.
\end{CodeOutput}
\begin{CodeInput}
R> summary(spi.ergm.all)
\end{CodeInput}
\begin{CodeOutput}
==========================
Summary of model fit
==========================

Formula:   spi ~ edges + nodematch("Assembly") + triangle("Assembly") +
 gwdegree(0.5, fixed = T) + graphletCount(c(6, 7, 8))

Iterations:  20

Monte Carlo MLE Results:
                   Estimate Std. Error MCMC %  p-value
edges              -4.80106    0.73658      8  < 1e-04 ***
nodematch.Assembly  2.11636    0.66232      5 0.001428 **
triangle.Assembly   3.27864    0.53805      0  < 1e-04 ***
gwdegree            1.12902    1.21795      1 0.354095
graphlet.6.Count   -0.12037    0.04122      2 0.003560 **
graphlet.7.Count   -0.46225    0.16905      0 0.006330 **
graphlet.8.Count   -2.31074    0.68949      0 0.000826 ***
---
Signif. codes:  0 '***' 0.001 '**' 0.01 '*' 0.05 '.' 0.1 ' ' 1

     Null Deviance: 1910.3  on 1378  degrees of freedom
 Residual Deviance:  596.7  on 1371  degrees of freedom

AIC: 610.7    BIC: 647.3    (Smaller is better.)
\end{CodeOutput}
\end{CodeChunk}

Once again we find positive, significant effects for homophily within complexes and triadic closure within complexes.  Controlling for this, we find negative, significant effects for graphlet terms $G_6$, $G_7$, and $G_8$.  

\begin{figure}[!h]
  \centering
  \includegraphics{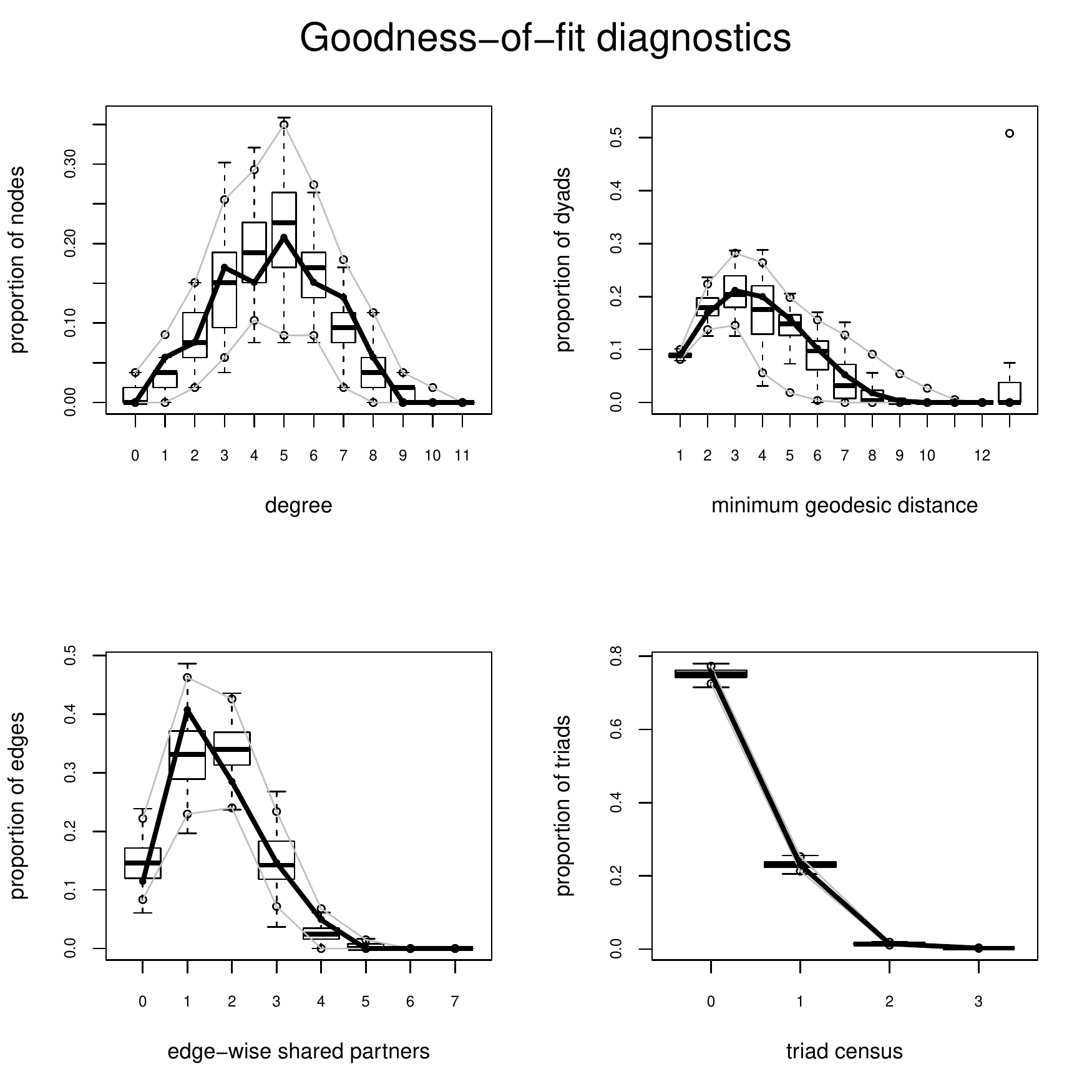}
  \caption{The solid black line in each plot represents the protein network's observed statistics.  The boxplots illustrate the statistics for our simulated networks, as produced by the MLE.}
  \label{f_SPIgof}
\end{figure}

Our final model appears to have converged without any notable issues \citep{modelChecking}.  We now assess the goodness of fit.  As Figure~\ref{f_SPIgof} indicates, our model closely approximates the observed network; our simulated networks show no clear deviations from the observed statistics on degree, geodesic distance, shared partners, or the triad census.

\begin{figure}[!h]
  \centering
  \includegraphics[width=\textwidth]{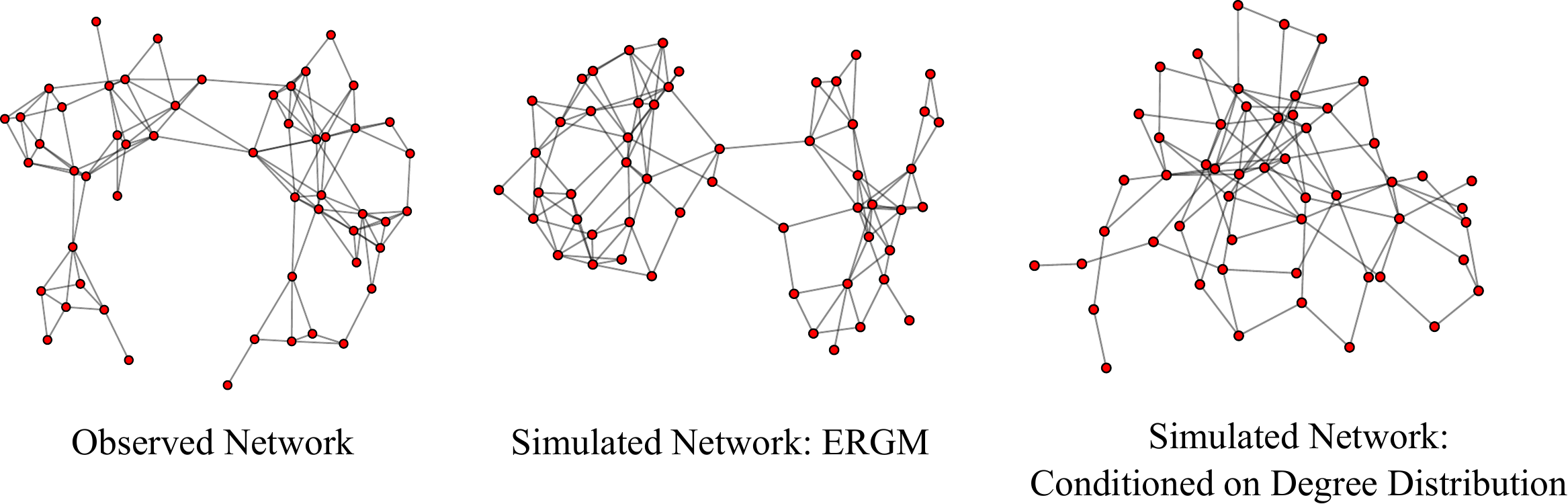}
  \caption{Observed protein network (left), typical protein network simulated by our final model (middle), and typical random graph produced by holding constant the observed network's degree distribution (right).}
  \label{f_comparison} 
\end{figure}

It is interesting to compare the results of our joint, multivariate analysis with the marginal tests conducted by \citet{MiloScience2004}.  \citet{MiloScience2004} find that the network overrepresents graphlets $G_6$, $G_7$, and $G_8$ and underrepresents $G_3$ and $G_4$.  After controlling for other factors (particularly clustering within each complex), we find no evidence of additional underrepresentation or overrepresentation of $G_3$ or $G_4$; further, we actually find that the network appears biased against formation of graphlets $G_6$, $G_7$, and $G_8$, once other terms are accounted for.  The discrepancy here is due to the use of marginal tests by \citet{MiloScience2004}.  To determine whether a graphlet occurs more or less often relative to chance, they compare the number of observed graphlets to the number observed in a set of random graphs conditioned on the degree distribution (a form of \emph{conditional uniform graph test}).  For this protein network, such random graphs bear little resemblance to the data in question (Figure~\ref{f_comparison}), and in particular do not include effects related to the fact that the structure is a composite of two distinct complexes.  While this does not make the results of such tests wrong per se, it does render them unable to distinguish between structural biases arising from simple features arising from the data collection process, and those arising from more subtle and informative biochemical mechanisms.  The marginal approach is also unable to unravel the joint influence of multiple biases simultaneously; because graphlet structures are dependent upon one another, over or underrepresentation of multiple graphlets (relative to a uniform baseline) may actually be the result of biases to a smaller number of features.  Such complexities are difficult to unravel using marginal tests, and are more flexibly handled via the ERGM framework.

Our analysis underscores the fact that one can obtain misleading conclusions when trying to use marginal tests to assess graphlet counts, particularly when the baseline distribution being employed does not incorporate extremely basic features of the system being studied.  While inference for complex, highly dependent systems is difficult under the best of conditions, the generative nature of the ERGM framework allows us to assess the adequacy of our models by comparison to features of the original data; given that we have identified a model that is both sensible and that successfully regenerates the important properties of the observed network, we have a stronger basis for subsequent investigation than would be obtained from simple rejection of a null hypothesis.

By using an ERGM approach and incorporating our graphlet terms, we are able to produce more sophisticated models of protein networks that include not only network motifs but also other important biological and/or chemical properties of the system in question.  Scholars in a variety of biological sub-disciplines have begun to use ERGMs to model many different types of networks, including protein-protein interaction networks \citep{Bulashevska2011, Clark2012}, neural networks \citep{so83003,Simpson2011,Simpson20121117}, and metabolic networks \citep{Saul01102007}.  Introducing the tools from the \pkg{ergm.graphlets} package to the network community should enhance the field's ability to model graphlet counts in the context of network motifs or any other application where one is interested in counts of small, undirected, induced subgraphs.

\section[Algorithms and implementation]{Algorithms and implementation}
\label{sec:algoritmic}
\begin{figure}[t]  
  \centering
  \includegraphics[width=\textwidth]{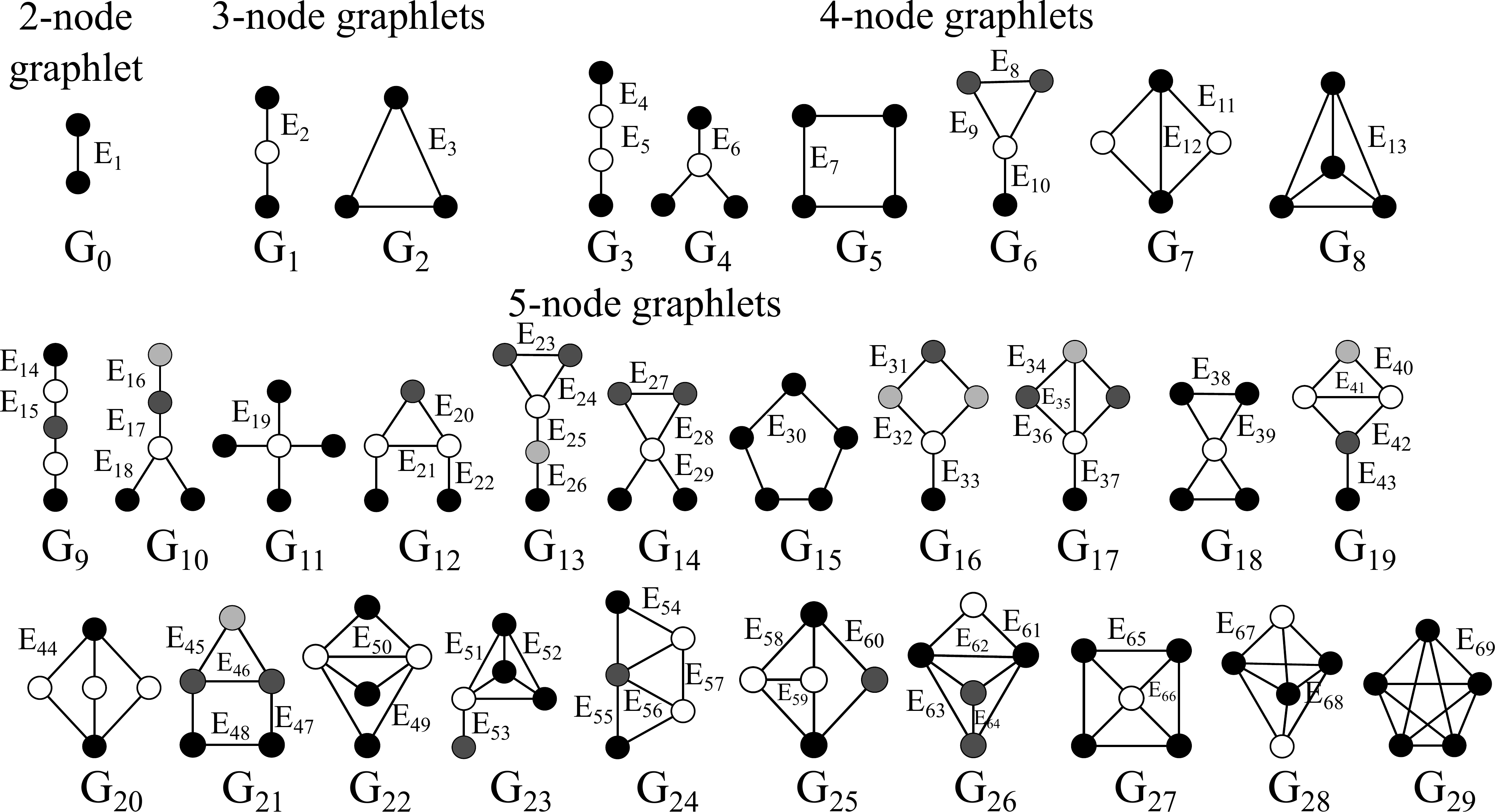}
  \caption{The edge automorphism orbits in 2-, 3-, 4- and 5-node graphlets. Adapted from \citet{edgeOrbits}.}
  \label{fig:edgeTypes}
\end{figure}

The terms in the \pkg{ergm.graphlets} package are implemented using the \pkg{ergm.userterms} package \citep{ergm.userterms}. The \pkg{ergm.userterms} package enables users to introduce new model terms into the \pkg{ergm} package by implementing \proglang{C} code which calculates the change statistics of the new term. For  the \pkg{ergm.graphlets} package, the change score function should answer the question: how do the graphlet counts in the network and graphlet degrees of the nodes change when an edge is flipped in the network? This question can be answered efficiently by \emph{touching} the graphlets on the flipped edge and counting only the graphlets that are going to be affected by the edge flip. For this purpose, we identify all \emph{edge automorphism orbits} in graphlets with 2, 3, 4 and 5 nodes. The 69 different edge automorphism orbits are in Figure~\ref{fig:edgeTypes} \citep{edgeOrbits}. In this section, we use \emph{node orbits} for graphlet orbits that are provided in Figure~\ref{fig:graphlets} and \emph{edge orbits} for edge automorphism orbits in Figure~\ref{fig:edgeTypes} for clarity.

We apply a brute-force search algorithm for computing the change score for graphlet terms. For each flipped edge, the edge orbits that are related with the queried graphlet structure are mapped on the flipped edge and the neighborhood of that edge is searched for nodes that complete the graphlet structure. For each node combination that completes the graphlet structure, the count of the affected graphlets is incremented by one. For identifying the change in the count of a specific graphlet, the computation is performed only for relevant edge orbits. The relations among graphlets and edge orbits are summarized in Table~\ref{tab:graphletEdgeRelations}. For example, the change score for the counts of graphlet $G_{11}$ and $G_{12}$ can be calculated by counting $E_{19}, E_{20}, E_{21}, E_{22}, E_{27}, E_{36}, E_{40}, E_{48}$. After counting these edge orbits, the change score for $G_{11}$ is equal to $(E_{19} - E_{27})$ and the change score for $G_{12}$ is equal to $(E_{20} + E_{21} + E_{22} - E_{36} - E_{40} - E_{48})$ where $E_{x}$ represents the number of graphlets counted by placing edge orbit $x$ on the flipped edge. By counting the graphlet change scores based on edge orbits, we do not only restrict the counting process to graphlets that are affected from the edge flip, but also avoid repeated counting of the same edge orbit for different graphlet counts. For instance, $E_{3}$ affects the count of $G_{2}$ positively and the count of $G_{1}$ negatively. With our implementation, the number of graphlets affected by $E_3$ is counted only once, and this change score is computed for identifying the changes in the counts of both $G_1$ and $G_2$. The edge orbit based counting procedure is applied for computing the change scores for all the terms in the \pkg{ergm.graphlets} package.

\begin{table}[!h]
\centering
\footnotesize
\begin{tabular}{|c|l|l||c|l|l|}
\hline
~ 		 & \multicolumn{2}{c||}{\textbf{Edge Automorphism}} & ~ & \multicolumn{2}{c|}{\textbf{Edge Automorphism}}\\
\textbf{Graphlet} & \textbf{Positive} & \textbf{Negative} & \textbf{Graphlet} & \textbf{Positive} & \textbf{Negative} \\
\hline
\hline
$\mathbf{G_{0}}$   & E$_{1}$ 							& -								& $\mathbf{G_{15}}$ & E$_{30}$				& E$_{46}$ 			\\
$\mathbf{G_{1}}$   & E$_{2}$ 							& E$_{3}$						& $\mathbf{G_{16}}$ & E$_{31}$, E$_{32}$, E$_{33}$		& E$_{35}$, E$_{41}$, E$_{44}$, E$_{45}$	\\
$\mathbf{G_{2}}$   & E$_{3}$ 							& -								& $\mathbf{G_{17}}$ & E$_{34}$, E$_{35}$, E$_{36}$, E$_{37}$	& E$_{49}$, E$_{52}$, E$_{54}$		\\
$\mathbf{G_{3}}$   & E$_{4}$, E$_{5}$ 					& E$_{7}$, E$_{9}$ 					& $\mathbf{G_{18}}$ & E$_{38}$, E$_{39}$			& E$_{57}$				\\
$\mathbf{G_{4}}$   & E$_{6}$							& E$_{8}$ 							& $\mathbf{G_{19}}$ & E$_{40}$, E$_{41}$, E$_{42}$, E$_{43}$	& E$_{51}$, E$_{55}$, E$_{60}$		\\
$\mathbf{G_{5}}$   & E$_{7}	$						& E$_{12}$							& $\mathbf{G_{20}}$ & E$_{44}$				& E$_{50}$, E$_{59}$			\\
$\mathbf{G_{6}}$   & E$_{8}$, E$_{9}$, E$_{10}$			& E$_{11}$							& $\mathbf{G_{21}}$ & E$_{45}$, E$_{46}$, E$_{47}$, E$_{48}$	& E$_{56}$, E$_{58}$ 		\\
$\mathbf{G_{7}}$   & E$_{11}$, E$_{12}$					& E$_{13}$							& $\mathbf{G_{22}}$ & E$_{49}$, E$_{50}$			& E$_{64}$				\\
$\mathbf{G_{8}}$   & E$_{13}$							& - 								& $\mathbf{G_{23}}$ & E$_{51}$, E$_{52}$, E$_{53}$ 	& E$_{61}$				\\
$\mathbf{G_{9}}$   & E$_{14}$, E$_{15}$					& E$_{21}$, E$_{24}$, E$_{30}$, E$_{32}$	& $\mathbf{G_{24}}$ & E$_{54}$, E$_{55}$, E$_{56}$, E$_{57}$	& E$_{63}$, E$_{65}$			\\
$\mathbf{G_{10}}$  & E$_{16}$, E$_{17}$, E$_{18}$			& E$_{20}$, E$_{23}$, E$_{28}$, E$_{31}$	& $\mathbf{G_{25}}$ & E$_{58}$, E$_{59}$, E$_{60}$		& E$_{62}$, E$_{66}$			\\
$\mathbf{G_{11}}$  & E$_{19}$							& E$_{27}$							& $\mathbf{G_{26}}$ & E$_{61}$, E$_{62}$, E$_{63}$, E$_{64}$	& E$_{67}	$		\\
$\mathbf{G_{12}}$  & E$_{20}$, E$_{21}$, E$_{22}$			& E$_{36}$, E$_{40}$, E$_{48}$			& $\mathbf{G_{27}}$ & E$_{65}$, E$_{66}$		& E$_{68}$			\\
$\mathbf{G_{13}}$  & E$_{23}$, E$_{24}$, E$_{25}$, E$_{26}$	& E$_{39}$, E$_{42}$, E$_{47}$			& $\mathbf{G_{28}}$ & E$_{67}$, E$_{68}$			& E$_{69}$			\\
$\mathbf{G_{14}}$  & E$_{27}$, E$_{28}$, E$_{29}$ 			& E$_{34}$, E$_{38}$					& $\mathbf{G_{29}}$ & E$_{69}$				& - 				\\
\hline
\end{tabular}

\caption{The relations between graphlet types and edge automorphism orbits. The ``Positive'' columns list the edge automorphism orbits that increase the graphlet count, and the ``Negative'' columns list the edge automorphism orbits that decrease the graphlet count when an edge is added.}
\label{tab:graphletEdgeRelations}
\end{table}

The computational complexity of this approach is dependent on the average degree (and therefore the density) of the modelled network. The \emph{average degree} of a network is defined as the average number of ties that a node has in the network. The \emph{density} of a network is defined as $\frac {|E|} {{|N| \choose 2}}$ where $|E|$ is the number of edges and $|N|$ is the number of nodes in the graph. In average case, the computational complexity of the change counting procedure is $O(d^{2})$ where $d$ represents the average degree of a node. The worst case scenario occurs when searching for graphlet $G_{9}$ in a clique. In this case, the computational complexity of the function is $O(n^{3})$ where $n$ is the number of nodes in the network. But this situation occurs very rarely as most real-world networks are sparse.

\begin{figure}[tb]  
  \centering
  \includegraphics[width=0.2\textwidth]{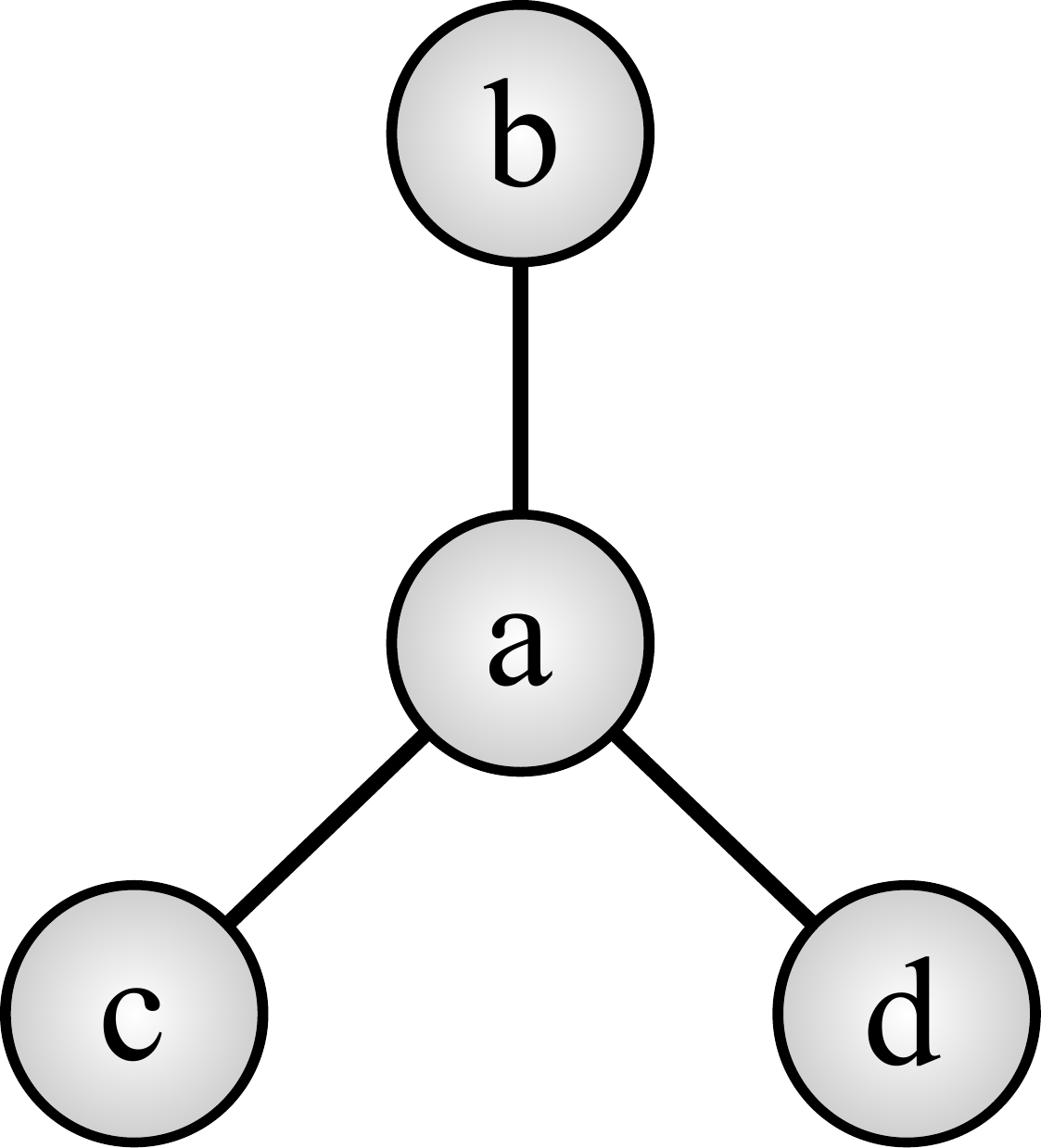}
  \caption{The subgraph that is used for illustrating the change statistics computation for the \code{grorbitCov} and \code{grorbitFac} terms.}
  \label{fig:subgraph}
\end{figure}

The four terms in the \pkg{ergm.graphlets} package are all implemented using edge orbits. However, the computation of the change scores differ slightly from each other depending on the way that the graphlet counts contribute to change statistics for these terms. The computation of the four terms in the \pkg{ergm.graphlets} package are explained as follows:

\begin{enumerate}
\item{\code{graphletCount(g): } The counting procedure is based on identification of graphlets. Therefore, each identified graphlet directly increments (or decrements) the change score for the related graphlet by 1. The change score for this term is computed by counting all edge orbits that are associated with the graphlets provided in argument $g$. When all required edge orbits are counted, these counts are summed to get the overall change in the number of graphlets. For example, the change score for graphlet $G_{12}$ is equal to the summation of $(E_{20} + E_{21} + E_{22} - E_{36} - E_{40} - E_{48})$ where $E_{x}$ represents the number of graphlets that touch the flipped edge on edge orbit $x$.
}

\item{\code{grorbitCov(attrname, grorbit): } This term relates a numeric nodal attribute with the graphlet degrees of the nodes according to Equation~\ref{eq:gorCov} as explained in Section~\ref{sec:terms}. The change score of this term depends on the graphlet degrees. Therefore, for each identified graphlet, the nodes of this graphlet are associated with the node orbits that they correspond to. Let us say that a graphlet of type $G_{4}$ is identified for the subgraph of nodes $a, b, c, d$, when the edge ($a$, $b$) is added into network. The identified subgraph is in Figure~\ref{fig:subgraph}. Then the change score for node orbit 6 is incremented by $X_{b} + X_{c} + X_{d}$, and the change score for node orbit 7 is incremented by $X_{a}$, where $X$ is the attribute vector keeping the attribute values for all nodes. The same logic applies when an edge is removed from the network. The final change score is obtained by summing these values for all edge orbits that are related with the graphlet that the query node orbit belongs to.
}

\item{\code{grorbitFactor(attrname, grorbit, base): } This term relates a categorical attribute with the graphlet degrees of the nodes according to Equation~\ref{eq:gorFactor} as explained in Section~\ref{sec:terms}. The change score of this term depends on the graphlet degrees. When the flip of an edge affects a node orbit, the change score that relates the category of the affected node with the node orbit is incremented (or decremented) by 1. Let us say a graphlet of type $G_{4}$ is identified for the subgraph of nodes $a, b, c, d$, when an edge ($a$, $b$) is added into the network. The identified subgraph is in Figure~\ref{fig:subgraph}. Node $a$ and $b$ belong to ``Category 1'', $c$ and $d$ belong to ``Category 2''. In this scenario, the change score for ``Node Orbit 7, Category 1'' and ``Node Orbit 6, Category 1'' will increase by 1 with the contribution of nodes $a$ and $b$. The change score for ``Node Orbit 6, Category 2'' will increase by 2 because of the nodes $c$ and $d$. The same logic applies  when an edge is removed from the network. The final change score is obtained by summing these values for all edge orbits that are related with the graphlet that the query node orbit belongs to.
}

\item{\code{grorbitDist(grorbit, d): } This term identifies the change in the graphlet degree distribution of a node orbit when an edge is flipped during the MCMC process, as explained in Section~\ref{sec:terms}. The change score computation for this term is slightly different from the other terms, as graphlet degrees for all nodes in the network are required for the computation. In order to reduce the computational complexity of the problem, we compute the graphlet signatures of all nodes at the beginning of the MCMC procedure. During the execution of the MCMC procedure, we update these signatures using the change scores. The computation of the changes in graphlet degrees of the nodes is performed similar to the algorithm applied for the other terms. However, as graphlets can convert to each other with the addition or removal of edges, the counting procedure should be applied for all edge orbits. Therefore, it is not possible to restrict the counting procedure to edge orbits that are related with the query node orbits. For these reasons, the computational complexity of this term is higher than the other terms in the \pkg{ergm.graphlets} package. We implement \code{grorbitDist} term only for graphlets with 2, 3, and 4 nodes, because of the high computational complexity of the computation of change score for graphlets with 5 nodes.
}
\end{enumerate}

We first test the correctness of the terms in the \pkg{ergm.graphlets} by using the \code{summary} function. As explained in Section~\ref{sec:ergm}, the \code{summary} function computes the statistics for the provided terms. If the change score function is implemented correctly, the \code{summary} function produces the actual value of the term statistic for the provided network. In this respect, we validated the correctness of the \code{graphletCount} term by running:
\begin{CodeChunk}
\begin{CodeInput}
R> summary(ntwk ~ graphletCount)
\end{CodeInput}
\end{CodeChunk}
The output of the \code{summary} function was exactly the same with the graphlet counts produced by the graphlet counting implementation of \citet{graphlets2}. Evaluating the correctness of the \code{grorbitCov} term is slightly different from \code{graphletCount} as it is related with a nodal attribute value. To test the correctness of this term, we first created a dummy node attribute that is named ``dummy''. This node attribute has value 1 for all nodes in the network. We validated the correctness of \code{gorbitCov} by running: 
\begin{CodeChunk}
\begin{CodeInput}
R> summary(ntwk ~ grorbitCov("dummy"))
\end{CodeInput}
\end{CodeChunk}
The output of the \code{summary} function was exactly the same with the sum of the graphlet degrees of all nodes for all node orbits. We repeated this test with weighted attribute values, e.g., when all attribute values are set to 2. The correctness of the results are also validated for this case. The validation for the \code{grorbitFactor} is similar to \code{grorbitCov} term. We assigned the same value for the category attribute, named ``dummy'',  of all nodes and called the \emph{summary} function as:
\begin{CodeChunk}
\begin{CodeInput}
R> summary(ntwk ~ grorbitFactor("dummy", 0:72, 0))
\end{CodeInput}
\end{CodeChunk}
This call correctly returns the sum of graphlet degrees of all terms. When the category value is changed with another value, the output of the call does not change. Finally, for validating the \code{grorbitDist} term, we called the \code{summary} function as:
\begin{CodeChunk}
\begin{CodeInput}
R> summary(ntwk ~ grorbitDist(0:14 , 0:10))
\end{CodeInput}
\end{CodeChunk}
This call produces the correct graphlet degree distributions for all node orbits as validated in comparison with the output of the implementation of \citet{graphlets2}.

We also validated the correctness of our implementation by performing simulations on ERGMs that contain graphlet terms. In these tests, we defined ERGMs containing an \code{edge} term and a graphlet term. We manually set the model coefficient for the graphlet related term to various positive and negative values. We simulated 30 networks from each of these ERGM models. With these simulations, we validated that positive coefficients promote the count of the related graphlet in the simulated networks. The count of related graphlet increase up to a certain coefficient value. After this threshold, the simulated networks contain the maximum possible number of related graphlets in the simulated networks. Similarly, negative coefficients have an effect of suppressing the appearance of the graphlet in the simulated networks. As the coefficient value gets closer to 0, the suppressing effect disappears. The range that the graphlet counts increase with the coefficient depends on the coefficients of the other terms in the ERGM model.

\section[Discussion]{Discussion}
\label{sec:discussion}
The \pkg{ergm.graphlets} package introduces four new terms into the \pkg{ergm} package which enable ERG modeling using the graphlet properties of a network. \code{graphletCount} term enables defining ERGMs based on the number of graphlets in the network. \code{grorbitCov} term uses the relation between a numeric nodal attribute with a specific topological structure in order to introduce the nodal attribute relations into the model. \code{grorbitFactor} term is similar to \code{grorbitCov} term except that it relates categorical attributes with graphlet degrees. \code{grorbitDist} term uses the graphlet degree distribution for ERG modeling. The \code{graphletCount}, \code{grorbitCov} and \code{grorbitFactor} terms are defined for graphlets with 2, 3, 4 and 5 nodes. Because of the computational complexity issues, \code{grorbitDist} is not defined for 5 node graphlets.

The \pkg{ergm} package already contains a number of terms that are related with the subgraph properties of networks, e.g., \code{triangle}, \code{twopath}, \code{threepath}, \code{cycle}, \code{kstar}. \pkg{ergm.graphlets} expands on these capabilities by providing a complete catalogue of small induced subgraphs for use in ERG modeling.  Induced subgraph counts differ from partial subgraph counts, with the majority of existing terms (e.g., those named above) being of the latter type.  For example, an isolated complete subgraph of order 3 simultaneously contributes 3 two-paths and one triangle, but does not contribute any copies of graphlet $G_1$ (because $G_1$ counts only \emph{open} two-paths).  As both types of subgraph counts are potentially useful, \pkg{ergm.graphlets} expands the options available to the ERGM modeler. 

Model degeneracy, instability, and sensitivity are currently important challenges for modeling within the ERGM framework \citep{modelDegeneracy, instability}. For some combinations of model terms, the MCMC procedure may fail to converge within a reasonable number of iterations: this is generally because the graph distribution associated with the specified model family are ill-behaved. Like most dependence terms, the terms in the \pkg{ergm.graphlets} package sometimes suffer from these instability issues, depending on the modelled network and the other terms in the ERGM. Typically, degeneracy problems are currently handled either by using user-selected terms whose effects partially cancel (e.g., using sparse graphlets and complete graphlets together) or using curved exponential family models \citep{curvedExponential, Bernoulli, instability} that systematically combine large numbers of terms in a manner that balances their total effect. The former technique requires having an intuition about the structure of the data and a number of trials with different combinations of terms under this intuition. It can be hard to identify the best terms for generating an ERGM model and there is currently no general solution that works well in all settings. Our experience suggests that graphlet terms for which the change score is non-zero for most of the steps in the MCMC procedure are good terms with which to start the modeling process. For example, it is not reasonable to model a sparse network using clique-like graphlets, as the change score will be 0 for most of the MCMC steps. In this respect, the graphlet terms that are expected to be overrepresented in the network can also be good candidate terms to start ERGM modeling. Using terms of the same graphlet size together usually improves the convergence of the MCMC process, since smaller graphlets might already be contained in a number of larger graphlets and this causes dependency issues among model terms. We have also observed that MCMC procedure converges faster when graphlets containing closed-loop structures (e.g., triangles, cycles) are excluded from the model definition: This is mainly because of the instability of these terms, as explained in \citet{instability}. As more data sets are subjected to analysis using ERGMs (and models with graphlet terms in particular), better heuristics are likely to emerge.  

Past work with (non-induced) subgraph terms has suggested that curved exponential family models can also be used for improving degeneracy issues. In curved exponential families, the parameters associated with model statistics are constrained to lie on a non-linear surface of reduced dimension, forcing them to remain in a fixed relationship with one another; this can be helpful when dealing with intrinsically correlated graph statistics, as very precise weighting may be needed to avoid the degenerate regime.  Examples of curved terms include the \code{gwdegree}, \code{gwdsp}, and \code{gwesp} terms of the \pkg{ergm} package, as well as the closely related \emph{alternating k-star} and \emph{alternating path} statistics of \citet{snijders.et.al:sm:2006}.  Because graphlet statistics do not ``nest'' in the same way as non-induced subgraph statistics, they may benefit from novel formal development.  On the other hand, some ideas used in existing curved families---e.g., geometrically weighted degree distributions---could potentially be applied to graphlet orbit degrees in a relatively straightforward manner.  This would seem to be a promising direction for future research.

When the over or underrepresentation of a specific graphlet statistic is of particular interest but inclusion of this statistic into one's model proves difficult, another alternative is the use of a simplified model omitting the statistic as a reference distribution against which to test the observed graphlet statistic.  Specifically, let $t'$ be the statistic of interest, and let $t$ be the vector of statistics in the best-fitting model without $t'$.  A test of the hypothesis that the parameter $\theta'$ associated with the joint model $(t'\cup t, \theta' \cup \theta)$ is non-zero can be conducted by examining the quantiles of $t'(y)$ in the distribution of $t(Y)$, where $Y \sim \mathrm{ERG}(\hat{\theta},t)$ and $\hat{\theta}$ is the MLE of $\theta$ given $y$.  This approach (which was one motivation for the original development of ERGMs) is described in more detail by \citet{holland.leinhardt:jasa:1981}.

Although we have tried to minimize the complexity of the change score computation, there is still room for improving the graphlet counting process. We apply a brute-force algorithm, which tries to minimize the number of computations: this gives an exact solution. Further gains in efficiency may be possible. These improvements would enable the implementation of \code{grorbitDist} for graphlets with 5 nodes. The model coefficients for terms related with larger graphlets would also be estimated more quickly with these improvements.

In addition to their inferential value, we note that the terms in the \pkg{ergm.graphlets} package can be used for evaluating the goodness-of-fit of an ERGM model estimate based on other (non-graphlet terms). When a model (with or without graphlet related terms) is estimated, the quality of this model in explaining the structure of the data in terms of graphlet properties of the network can be assessed by simulating new networks from the model and using \code{summary} function to compute the graphlet counts and graphlet degree distributions. The graphlet properties of the network can be compared with these simulation results to evaluate whether the topology of the network fits to the topology described by the model. An example that describe how this test can be performed is explained in \citet{statnetTutorial}.

In conclusion, the \pkg{ergm.graphlets} package extends the functionality of the \pkg{ergm} package by incorporating support for counts of small induced subgraphs, orbit membership, and related quantities.  The introduced terms are of particular utility when modeling processes such as brokerage, functional mediation, or other phenomena that depend not only on the edges that are present within a graph, but also on those that are absent.  Such processes are common in both social and biological systems, and the ability to capture them is an important goal of modern network analysis.

\section*{Acknowledgments} 
We thank Kai Sun and Miles Mulholland for helpful suggestions and comments about the manuscript. The project was supported by ERC Starting Independent Researcher Grant 278212, NSF CDI OIA - 1028394 grant, ARRS project J1-5454, and the Serbian Ministry of Education and Science Project III44006.

\bibliography{references}{}

\end{document}